\documentclass{aa}
\usepackage{graphicx}
\usepackage{natbib}
\usepackage{txfonts}
\bibpunct{(}{)}{;}{a}{}{,}
\newcommand{\ha}{H$\alpha$}
\newcommand{\hb}{H$\beta$}

\newcommand{\kms}{\,km\,s$^{-1}$}
\newcommand{\myr}{\,$M_{\sun}\,{\rm yr}^{-1}$}
\newcommand{\ro}{\,$R_{\sun}$}

\newcommand{\lo}{\,$L_{\sun}$}
\newcommand{\kpc}{{\rm kpc}}
\newcommand{\cmt}{\,cm$^{-3}$}

\newcommand{\ecs}{$\rm\,erg\,cm^{-2}\,s^{-1}$}
\newcommand{\rec}{$\rm\,erg\,cm^{3}\,s^{-1}$}
\newcommand{\ecsa}{$\rm\,erg\,cm^{-2}\,s^{-1}\,\AA^{-1}$}
\begin{document}
\title{Formation of a disk structure in the symbiotic binary AX~Per \\
       during its 2007-10 precursor-type activity\thanks{Tables 2, 3 
       and 5 are available at http://www.aanda.org}}

\author{A.~Skopal\thanks{Visiting Astronomer: Asiago Astrophysical 
                         Observatory}
        \inst{1}
        \and
        T.~N.~Tarasova
        \inst{2}
        \and
        Z.~Carikov\'a
        \inst{1}
        \and
        F.~Castellani
        \inst{3}
        \and
        G.~Cherini
        \inst{3}
        \and
        S.~Dallaporta
        \inst{3}
        \and
        A.~Frigo
        \inst{3}
        \and
        C.~Marangoni
        \inst{3}
        \and
        S.~Moretti
        \inst{3}
        \and
        U.~Munari
        \inst{4}
        \and
        G.~L.~Righetti
        \inst{3}
        \and
	A.~Siviero
        \inst{5}
        \and
        S.~Tomaselli
        \inst{3}
        \and
        A.~Vagnozzi
        \inst{3}
        \and
        P.~Valisa
        \inst{3}
        }
\institute{Astronomical Institute, Slovak Academy of Sciences,
           059\,60 Tatransk\'{a} Lomnica, Slovakia 
       \and
          Crimean Astrophysical Observatory, Nauchny, Ukraine
       \and
          ANS Collaboration, c/o Osservatorio Astronomico di Padova, 
          Sede di Asiago, I-36032 Asiago (VI), Italy
       \and
          INAF Osservatorio Astronomico di Padova, Sede di Asiago, 
          I-36032 Asiago (VI), Italy 
       \and
          Dipartimento di Astronomia, Universita' di Padova, Osservatorio
          Astrofisico, I-36012 Asiago (VI), Italy
         }
\date{Received / Accepted }

\abstract
 {
  AX~Per is an eclipsing symbiotic binary. During active phases, 
  deep narrow minima are observed in its light curve, and 
  the ionization structure in the binary changes significantly. 
  From $\sim$2007.5, AX~Per entered a new active phase. 
 }
 {
  We aim to derive the ionization structure in the binary and 
  its changes during the recent active phase. 
 }
 {
  We used optical high- and low-resolution spectroscopy and 
  $UBVR_{\rm C}I_{\rm C}$ photometry. 
  We modeled the SED in the optical and broad wings of the 
  \ha\ line profile during the 2007-10 higher level of 
  the AX~Per activity. 
 }
 {
After 10 orbital cycles ($\sim 18.6$ years), we again measured 
the eclipse of the hot component by its giant companion in the 
light curve. We derived a radius of $27 \pm 2$\ro\ 
for the eclipsed object and $115 \pm 2$\ro\ for the eclipsing 
cool giant. 
The new active phase was connected with a significant enhancement of 
the hot star wind. From quiescence to activity, the mass-loss rate 
increased from $\sim 9\times 10^{-8}$ to $\sim 3\times 10^{-6}$\myr,
respectively. The wind causes the emission of the He$^{++}$ zone, 
located in the vicinity of the hot star, and also is the reason 
for the fraction of the [\ion{O}{iii}] zone at farther distances. 
Simultaneously, we identified a variable optically thick warm 
($T_{\rm eff} \sim 6000$\,K) source that contributes markedly to 
the composite spectrum. The source was located at the hot star's 
equator and has the form of a flared disk, whose outer rim 
simulates the warm photosphere. 
 }
 {
The formation of the neutral disk-like zone around the accretor 
during the active phase was connected with its enhanced wind. It 
is probable that this connection represents a common origin of 
the warm pseudophotospheres that are indicated during the active 
phases of symbiotic stars. 
 }
\maketitle
\section{Introduction}

Symbiotic stars are interacting binary systems comprising 
a cool giant and a hot compact star, mostly a white dwarf (WD), 
on, typically, a few years' orbit. The WD accretes from the 
giant's wind, heats up to 1--2$\times 10^5$\,K, and becomes 
as luminous as $10^2 - 10^4$\lo. It ionizes the circumbinary 
environment and causes nebular emission. As a result, the 
spectrum of symbiotic stars consists of three basic components 
of radiation -- two stellar and one nebular. 
If the processes of the mass loss, accretion and ionization 
are in a mutual equilibrium, the symbiotic system releases 
its energy approximately at a constant rate and spectral 
energy distribution (SED). This stage is called the 
{\em quiescent phase}. Once this equilibrium is disturbed, 
the symbiotic system changes its radiation significantly, 
brightens in the optical by a few magnitudes and usually shows 
signatures of a mass outflow for a few months to years. 
We name this stage the {\em active phase}. Many particular 
aspects of this general view have been originally pointed out 
by, e.g., \cite{f+v82}, \cite{stb}, \cite{kw84}, \cite{nv87}, 
and are more recently discussed in \cite{cmm03}. 

To explain the high luminosity of hot components 
in symbiotic binaries, it was suggested that they are 
powered by stable hydrogen nuclear burning on the WD surface, 
which requires a certain range of accretion rates 
\citep[e.g.][]{paczyt78,fuji82}. 
During active phases, we observe a significant cooling of 
the spectrum produced by the hot star 
\citep[e.g.][ and references therein]{kw84}. 
The event of outbursts could result from an increase in the 
accretion rate above that which sustains the stable burning, 
which leads to an expansion of the burning envelope to an A--F 
type (pseudo)photosphere \citep[e.g.][]{tutyan76,pacrud80}. 
As a result, the pseudophotosphere will radiate at a lower 
temperature, and consequently shift the maximum of its SED 
from shorter 
to longer wavelengths, causing a brightening in the optical. 
Based on optical observations, this scenario was supported by 
many authors. Recently, \cite{siv+09} applied it to the 2008-09 
active phase of CI~Cyg. 
However, modeling the UV/near-IR continuum of symbiotic binaries 
with a high orbital inclination during active phases indicated 
an increase of both the stellar radiation from a warm 
(1--2$\times 10^{4}$\,K) pseudophotosphere and the nebular 
radiation within the optical wavelengths. 
This led to the suggestion that there is an edge-on disk 
around the accretor, whose outer flared rim represents 
the warm pseudophotosphere, and the nebula is placed 
above/below the disk \citep[][]{sk05}. 
In this contribution we test this scenario on the recent 
active phase of the symbiotic binary AX~Per. 

At present, AX~Per is known as an eclipsing symbiotic binary
\citep[][]{sk91} comprising a M4.5\,III giant \citep[][]{m+s99} 
and a WD on  a 680-d orbit \citep[e.g.][]{mk92,fekel+00}. 
The last active phase began in 1988, when AX~Per brightened
by $\sim$3\,mag in the visual, developed a specific 
phase-dependent modulation in the light curve (LC), and showed 
narrow eclipses at the position of the inferior conjunction of 
the giant \citep[e.g.][]{ivison+93}. Transition to quiescence 
happened during 1995, when the LC profile turned to the 
wave-like orbitally related variation at a low level 
of the star's brightness \citep[][ Fig.~1 here]{sk+01}. 
Recently, \cite{mu+09a} reported on a rapid increase in the 
AX~Per brightness by $\sim$1\,mag in $B$ during 2009 March. 
They pointed a similarity 
between this brightening phase and the short-duration flare 
that occurred in the AX~Per LC about one year before its 
major 1988-1995 active phase. Recently, \cite{mu+10} reported 
on a rise in brightness of AX~Per during 2010 November by 
$\sim$0.7\,mag in $B$, which supported their previous 
suggestion that the 2007-10 active phase was a precursor 
to an oncoming major outburst. 
From this point of view, understanding the precursor-type 
activity can aid us in understanding the nature of the 
Z~And-type outbursts better. 

Accordingly, in this paper we investigate the structure of the 
hot component in the binary (i.e. the hot star with the warm 
pseudophotosphere and the nebula) that developed during the 
2007-10 brighter phase of AX~Per. To achieve this aim, we 
analyze our multicolour photometric observations and 
the low- and high-resolution optical spectra. 
In Sect.~2 we summarize and describe our observations and data 
reduction. Section~3 describes our analysis and presents the 
results. Their discussion and conclusions are found in 
Sects.~4 and 5, respectively. 
%
%
%
\begin{figure*}
\centering
\begin{center}
\resizebox{\hsize}{!}{\includegraphics[angle=-90]{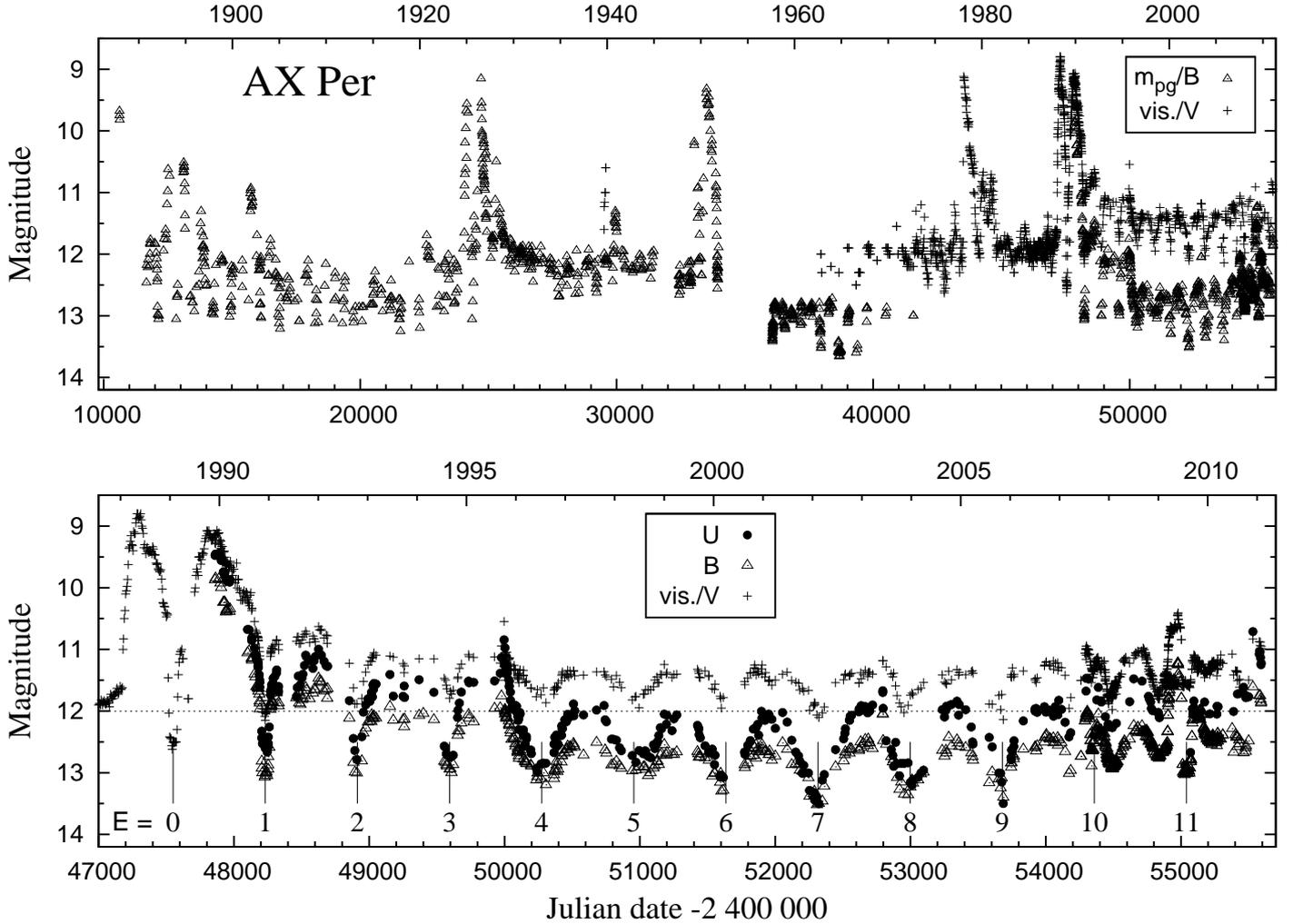}}
\caption[]{Top panel displays the historical LC of AX~Per from 
$\sim$1890 \citep[see Fig.~1 of][ and references therein]{sk+01}. 
We used visual magnitude estimates from the AAVSO International 
Database and those gathered by members of the AFOEV, which are 
available on CDS. Visual data were smoothed within 20-day bins. 
Bottom panel shows the $UBV$ LCs of AX~Per from its last 1988 
major outburst. 
Minima timing corresponds to the ephemeris given by Eq.~(1). 
          }
\end{center}
\end{figure*}

\section{Observations and data reduction}

Our observations of AX~Per during its 2007-10 higher level 
of activity were carried out at different observatories. 

Spectroscopic observations were secured with five different instruments: 
(1) 
the 1.22-m telescope, operated in Asiago by the Department of Astronomy 
of the University of Padova, was used with a B\&C spectrograph, 300 
ln/mm grating and ANDOR iDus 440A CCD camera, equipped with 
an EEV 42-10BU back illuminated chip (2048$\times$512 pixels of 
13.5~$\mu$m size); 
(2) 
the 1.82-m telescope, operated in Asiago by INAF Astronomical 
Observatory of Padova, mounting a {\small REOSC} echelle 
spectrograph equipped with an {\small AIMO E2VCCD47-10} 
back-illuminated CCD detector (1100$\times$1100 pixels of 
13\,$\mu$m size); 
(3) 
the 0.60-m telescope of the Schiaparelli Observatory in Varese, 
mounting a multi-mode spectrograph, able to provide both single 
dispersion, low-resolution spectra and Echelle spectra. 
The detector was an SBIG ST-10XME camera (2194$\times$51472 pixels 
of 6.8~$\mu$m size); 
(4) 
the 1.88-m telescope of the David Dunlap Observatory (DDO),
University of Toronto (DDO), equipped with a single-dispersion
spectrograph equipped with a Jobin Yovon Horiba CCD detector
(2048$\times$512 pixels of 13.5\,$\mu$m size); and finally 
(5) 
the 2.6-m Shajn telescope, operated by the Crimean Astrophysical 
Observatory (CrAO) in Nauchny, mounting a {\small SPEM} 
spectrograph in the Nasmith focus. The detector was a SPEC-10 CCD 
camera (1340$\times$100 pixel). 
Table~1 provides a journal of the spectroscopic observations. 
At each telescope the observations were carried out as multiple 
exposures to avoid saturation of the strongest 
emission lines. Various spectrophotometric standards were observed 
each night to flux into absolute units the spectra of AX~Per. 
The accuracy of the flux scale was checked against the photometric 
observations by integration
over the fluxed spectra of the $UBVR_{\rm C}I_{\rm C}$ photometric    
pass-bands. Correction for flat-, bias- and dark frames was carried 
out in a standard way with IRAF, as well subtraction of 
the background sky and scattered light. 
%
%
%
\begin{table}
\caption[]{Log of spectroscopic observations}
\begin{center}
\begin{tabular}{@{}c@{~~}c@{~~}c@{~~}c@{~~}c@{~~}l@{}}
\hline
\hline
 Date      & Julian date & Res. & Disp.     & $\lambda$ range  & Telescope\\
           & +2450000    & Power& (\AA/pix) &  (\AA)           &          \\
\hline
2007/07/31 &  4312.876 &10\,000 &    &    \hb,~ \ha\     & DDO 1.88m      \\
2007/11/20 &  4425.529 &        &1.81&    3824- 7575     & CrAO 2.6m      \\
2007/12/03 &  4438.438 &        &2.30&    3230- 7770     & Asiago 1.22m   \\
2008/01/23 &  4488.506 &10\,000 &    &    \hb,~\ha\      & DDO 1.88m      \\
2008/07/07 &  4655.473 &        &1.81&    3774- 7574     & CrAO 2.6m      \\
2008/08/10 &  4689.450 &        &1.81&    3749- 7575     & CrAO 2.6m      \\
2008/09/24 &  4734.609 &        &1.81&    3749- 7575     & CrAO 2.6m      \\
2008/10/13 &  4753.638 &20\,000 &    &    3690- 7300     & Asiago 1.82m   \\
2008/10/23 &  4763.417 &        &1.81&    3349- 7149     & CrAO 2.6m      \\
2008/11/08 &  4778.393 &        &1.81&    3749- 7500     & CrAO 2.6m      \\
2008/12/08 &  4809.390 &        &2.30&    3250- 7830     & Asiago 1.22m   \\
2008/12/21 &  4822.437 &20\,000 &    &    3690- 7300     & Asiago 1.82m   \\
2009/01/09 &  4841.426 &20\,000 &    &    3690- 7300     & Asiago 1.82m   \\
2009/03/24 &  4915.422 &        &2.12&    3890- 7930     & Varese 0.60m   \\
2009/04/01 &  4923.323 &        &2.30&    4200- 7200     & Asiago 1.22m   \\
2009/04/07 &  4929.303 &        &2.12&    3900- 8120     & Varese 0.60m   \\
2009/04/08 &  4930.301 &        &2.30&    3340- 7620     & Asiago 1.22m   \\
2009/04/09 &  4931.262 &        &1.81&    3600- 7575     & CrAO 2.6m      \\
2009/04/14 &  4936.281 &        &2.30&    3560- 7580     & Asiago 1.22m   \\
2009/04/14 &  4936.338 &17\,000 &    &    4030- 7510     & Varese 0.60m   \\
2009/05/05 &  4957.458 &        &2.12&    3850- 8270     & Varese 0.60m   \\
2009/08/05 &  5049.535 &        &2.30&    3320- 7660     & Asiago 1.22m   \\
2009/08/26 &  5070.565 &        &1.81&    3324- 7493     & CrAO 2.6m      \\
2009/09/24 &  5099.584 &        &1.81&    3350- 7500     & CrAO 2.6m      \\
2009/11/10 &  5146.485 &20\,000 &    &    3690- 7300     & Asiago 1.82m   \\
2010/08/07 &  5416.368 &        &2.12&    3910- 8540     & Varese 0.60m   \\
2010/08/16 &  5425.564 &        &1.81&    3358- 7500     & CrAO 2.6m      \\
2010/08/21 &  5430.348 &17\,000 &    &    3950- 8640     & Varese 0.60m   \\
2010/09/14 &  5454.560 &        &1.81&    3274- 7574     & CrAO 2.6m      \\
\hline
\end{tabular}
\end{center} 
\end{table}  

Multicolour $BVR_{\rm C}I_{\rm C}$ CCD photometry was obtained
with various telescopes operated by the ANS (Asiago Novae and 
Symbiotic Stars) Collaboration. Treatments for flat-, bias- and 
dark frames was carried out in a standard way. 
Photometric calibration and correction for color equations was 
carried out against the photometric sequence calibrated by 
\cite{h+m06} around AX~Per. The resulting data are presented 
in Table~2. 
Their uncertainties are the overall budget errors (which includes 
the Poissonian component and the transformation to the standard 
Landolt system). They are of a few $\times 0.01-0.001$\,mag. 

In addition, classical photoelectric $UBV$ measurements were 
performed by a single-channel photometer mounted in the Cassegrain 
focus of 0.6-m reflector at the Skalnat\'{e} Pleso observatory. 
Internal uncertainties of these one-day-mean measurements are 
of a few $\times 0.01$\,mag. These data are presented in Table~3. 

Arbitrary flux units of the high-resolution spectra were 
converted to absolute fluxes with the aid of the simultaneous 
$UBVR_{\rm C}I_{\rm C}$ photometry. To determine flux-points of 
the true continuum from the measured magnitudes, we calculated 
corrections for emission lines, $\Delta m_l$ \citep[see][]{sk07}, 
using our low-resolution spectra (Table~4). 
Then we interpolated these values to dates of the high-resolution 
spectra. Spectroscopic observations were dereddened with 
$E_{\rm B-V}$ = 0.27 \citep[][]{kw84} and the resulting parameters 
were scaled to a distance of 1.73\,kpc \citep[][]{sk00}. 
%
%
\begin{figure*}
\centering
\begin{center}
\resizebox{16cm}{!}{\includegraphics[angle=-90]{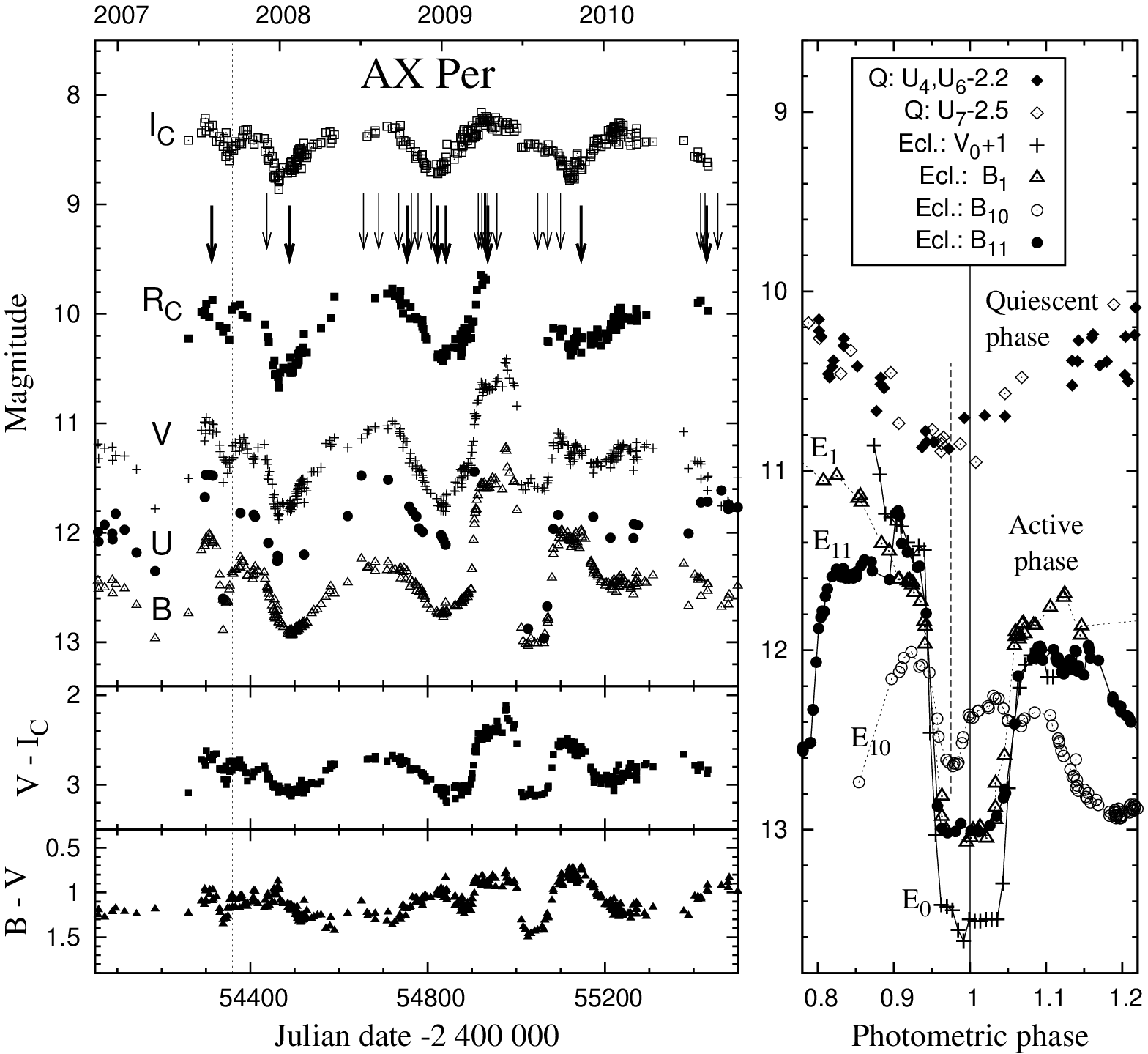}}
\caption[]{Left: Recent $UBVR_{\rm C}I_{\rm C}$ LCs 
of AX~Per covering its 2007-10 active phase. Thin and thick 
arrows denote the dates of our low- and high-resolution spectra, 
respectively (Table~1). Vertical dotted lines mark the time 
of eclipses according to Eq.~(1). 
Right: 
Minima profiles selected from LCs in Fig.~1, and folded with 
the ephemeris (1). Vertical dashed and solid line mark the light 
minima during quiescent and active phases, respectively. 
Denotation in keys, for example, "Q: U$_4$-2.2", means quiescent 
phase, U filter, epoch 4 and a shift by -2.2\,mag. 
          }
\end{center}
\end{figure*}
%
%
%
\addtocounter{table}{+2}
\begin{table}
\caption[]{Effect of emission lines on the 
           $UBVR_{\rm C}$ magnitudes}
\begin{center}
\begin{tabular}{ccccc}
\hline
\hline
Date &$\Delta U_l$ &$\Delta B_l$ &$\Delta V_l$ &$\Delta R_{{\rm C},l}$ \\
\hline
2007/11/20 &  --   & -0.26 & -0.07 & -0.20  \\
2007/12/03 & -0.14 & -0.24 & -0.05 & -0.15  \\
2008/07/07 & -0.10 & -0.30 & -0.06 & -0.20  \\
2008/08/10 & -0.11 & -0.33 & -0.07 & -0.22  \\
2008/09/24 & -0.12 & -0.34 & -0.07 & -0.24  \\
2008/10/23 & -0.20 & -0.35 & -0.08 & -0.25  \\
2008/11/08 &  --   & -0.34 & -0.07 & -0.24  \\
2008/12/08 & -0.18 & -0.32 & -0.06 & -0.21  \\
2009/03/24 &  --   & -0.19 & -0.05 & -0.13  \\
2009/04/01 &  --   & -0.15 & -0.06 & -0.12  \\
2009/04/07 &  --   & -0.17 & -0.05 & -0.12  \\
2009/04/08 &  --   & -0.20 & -0.05 & -0.12  \\
2009/04/09 &  --   & -0.21 & -0.06 & -0.14  \\
2009/04/14 &  --   & -0.19 & -0.06 & -0.12  \\
2009/05/05 &  --   & -0.14 & -0.04 & -0.10  \\
2009/08/05 & -0.15 & -0.23 & -0.06 & -0.10  \\
2009/08/26 & -0.10 & -0.21 & -0.07 & -0.13  \\
2009/09/24 &  --   & -0.15 & -0.06 & -0.13  \\
2010/08/07 &  --   & -0.30 & -0.07 & -0.22  \\
2010/08/16 & -0.18 & -0.36 & -0.08 & -0.27  \\
2010/09/14 & -0.22 & -0.39 & -0.10 & -0.31  \\
\hline
\end{tabular}
\end{center}
\end{table}    

\section{Analysis and results}

\subsection{Photometric evolution}

\subsubsection{Indications of a new active phase}

According to the LC evolution from 1988 (Fig.~1), 
the increase in the star's brightness by $\Delta U \sim$0.5\,mag 
from 2007 July signals that AX~Per entered a new active phase. 
An additional rapid brightening by $\Delta B \sim$0.8\,mag 
and a bluer index $B-V \sim 0.8$, observed during 2009 March 
\citep[][ and Fig.~2 here]{mu+09a}, supported the new activity 
of AX~Per. 
A significant change of the broad minima, observed during 
quiescent phases at/around the giant inferior conjunction, into 
a narrower and deeper eclipse, indicates the active phase of 
a symbiotic binary as well 
\citep[e.g.][ and Figs.~1 and 2 here]{bel79,bel91}. 
During the 2007 conjunction we measured a relatively small, 
V-type minimum, which position was shifted from the giant 
conjunction by $\sim -0.025\,P_{\rm orb}$. During the 
following conjunction the eclipse was deeper and broader, 
nearly rectangular in profile and placed at the 
inferior conjunction of the giant. The light observed 
during the totality became redder with $B-V \sim 1.4$ 
(see Fig.~2). 
Finally, we measured a light wave in the LC with a period 
of $\sim 0.5\,P_{\rm orb}$ and minima located around orbital 
phases 0.2 and 0.7 (Fig.~2, left). 
Its presence was transient, being 
connected only with the 2007-10 higher level of the star's 
activity (see Fig.~1). It can be seen in all filters, 
but especially in the $I_{\rm C}$ band is well pronounced 
also after the 2009 eclipse (see Fig.~2). 
This feature was observed in a symbiotic system for the first 
time. Its understanding requires a detailed analysis, which is, 
however, out of the scope of this paper. Therefore we postpone 
it to a future work. 

\subsubsection{Eclipses}

The minimum (eclipse), observed during the 2009 brightening, 
was very similar in profile to those observed during 
the major 1988-91 active phase. Its well pronounced profile 
closely covered by measurements, mainly in the $B$- and $V$-bands, 
allowed us to determine its contact times as 
  $T_1 = JD~2\,454\,998.50 \pm 0.79$, 
  $T_2 = JD~2\,455\,012.50 \pm 0.53$,
  $T_3 = JD~2\,455\,065.40 \pm 0.85$ and 
  $T_4 = JD~2\,455\,085.33 \pm 0.50$, 
where uncertainties include only those given by the data 
coverage, assuming a linear dependence of the star's brightness 
during the ingress to and ascent from the totality. 
We neglected other possible sources of errors. 
These times and their uncertainties indicate an asymmetric 
profile with $T_2 - T_1 = 14 \pm 1.3$\,d and 
             $T_4 - T_3 = 20 \pm 1.3$\,d, 
which implies an increase of the linear size of the eclipsed 
object during the totality. 
Therefore, we determined the middle of the eclipse 
as the average of all contact times, i.e. 
   $JD_{\rm Ecl.}(2009) = JD~2\,455\,040.43 \pm 0.67$. 
Using well-defined timings of other eclipses, 
   $JD_{\rm Ecl.}(1989) = JD~2\,447\,551.7 \pm 1.0$ 
and 
   $JD_{\rm Ecl.}(1990) = JD~2\,448\,231.6 \pm 0.6$ 
\citep[][]{sk91}, gives their ephemeris as 
\begin{equation}
  JD_{\rm Ecl.} = 2~447\,551.26(\pm 0.3) + 
                  680.83(\pm 0.11) \times E. 
\end{equation}
Our photometric ephemeris agrees within uncertainties with 
that given by the solution of the spectroscopic orbit, 
$T_{\rm sp. conj.} = 2~447\,553.3(\pm 5.6) +
                  682.1(\pm 1.4) \times E $, 
obtained from infrared radial velocities by \cite{fekel+00}. 
Finally, assuming the orbital inclination $i = 90^{\circ}$ 
\citep[see a summary in][]{sk+01} and 
the separation between the binary components, $a = 364$\ro\ 
\citep[from the mass ratio and the mass function published 
by][]{mk92,fekel+00}, the radius of the giant, 
   $R_{\rm g} = 115 \pm 2$\ro\
and that of the eclipsed object,
   $R_{\rm e} = 27 \pm 2$\ro. 
This result suggests an expansion of the giant radius 
by $\sim$13\ro, from $102 \pm 3$\ro, given 
by the contact times of the 1990 eclipse \citep[][]{sk94}. 
The asymmetry of the 2009 eclipse profile and the radius of 
$R_{\rm e} = 36 \pm 3$\ro\ derived from the 1990 eclipse 
profile suggest that the eclipsed object is subject 
to variation (see also Table~8 below). 
%
%
\begin{figure*}
\centering
\begin{center}
\resizebox{\hsize}{!}{\includegraphics[angle=-90]{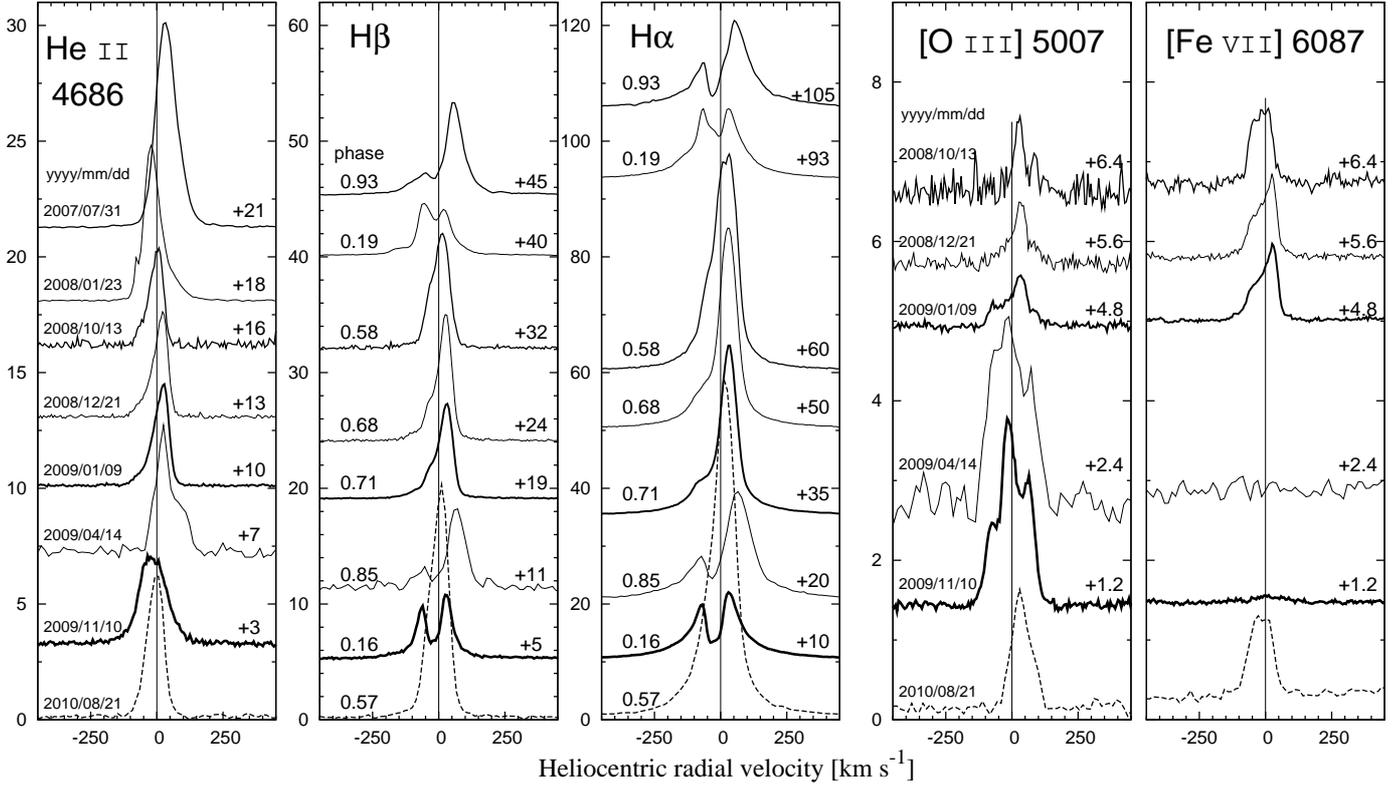}}
\caption[]{
Evolution in the selected line profiles along the 2007-10 active 
phase as observed in our high-resolution spectra. Fluxes are in units 
of $10^{-12}$\ecsa. Systemic velocity of --117.44\,\kms\ was 
subtracted. Small numbers at the right side of panels represent 
a shift of profiles with respect to the level of the local 
continuum. The spectra are dereddened. 
          }
\end{center}
\end{figure*}

\subsection{Spectroscopic evolution}

\subsubsection{The line spectrum}

During the investigated period of the 2007-10 activity, the 
optical spectrum of AX~Per was dominated by \ion{H}{i}, 
\ion{He}{i} and \ion{He}{ii} emission lines, although some 
faint emissions, produced by permitted transitions in 
ionized metals, as e.g. \ion{N}{iii}\,4641 and 
\ion{C}{iii}\,4647\,\AA, were also present. The spectrum of 
the forbidden lines was characterized mainly by 
nebular lines of [\ion{O}{iii}], [\ion{Ne}{iii}] and 
two [\ion{Fe}{vii}] lines at 5721 and 6087\,\AA. 
The Raman-scattered \ion{O}{vi}\,1032 line at 6825\,\AA\ 
was not present at all. 
Figure~3 shows evolution in the line profiles of 
\ion{He}{ii}\,4686\,\AA, \hb, \ha, [\ion{O}{iii}]\,5007\,\AA\ 
and [\ion{Fe}{vii}]\,6087\,\AA, as observed on our 
high-resolution spectra. 
Figure~4 demonstrates variations in the line fluxes along 
the active period, which we observed in all our spectra. Their 
quantities are introduced in Table~5 (available online). 
We also measured radial velocities (RVs) of main emission 
peaks in the line profiles of highly ionized elements (Table~6). 
Figure~5 displays their values compared to the radial velocity 
curves derived by other authors. 

{\sf Hydrogen lines:}
Profiles of hydrogen Balmer lines were significantly affected 
by a blue-shifted absorption component. Its presence can be 
recognized around the whole orbital cycle. From the orbital phase 
$\varphi = 0.85$ to $\varphi = 1.19$, it created a double-peaked 
profile (Fig.~3). 
In contrast, during the previous quiescent phase, the \ha\ 
line observed at similar orbital phases \citep[$\varphi = 0.81$ 
and $\varphi = 0.16$, as on 1998 Jan. 9 and Sept. 5, respectively, 
see Fig.~5 in][]{sk+01}, did not display the double-peaked 
profile. In addition, both the total flux and the extension
of the broad wings of mainly the \ha\ profile were significantly
higher than those measured during quiescent phase (Fig.~6 and
Sect.~3.2.2 below). It is important to note that during 
the major 1988-91 active phase of AX~Per a strong absorption cut 
the broad emission profile even at positions with the hot star 
in front \citep[see Figs.~5 and 8 of][]{ivison+93}. 
This indicates the presence of neutral material along the line 
of sight around the whole orbit, which attenuates line photons 
via the $b-b$ transitions. The strong difference 
in the line profiles measured during active and quiescent 
phases suggests that the density of the absorbing material 
enhances during active phases at the orbital plane. 
Finally, during the eclipse the total fluxes of \ha\ and \hb\ 
lines and those of their wings decreased by a factor of $ > 3$ 
with respect to their maximum values (Fig.~4, Table~5). The same 
behavior was observed during the 1994 eclipse 
\citep[see Fig.~6 of][]{sk+01}. Hence, during the active phases, 
most of the ionized region around the hot star is confined in 
a region that has about the size of the red giant's radius. 
%
%
\begin{figure}
\centering
\begin{center}
\resizebox{\hsize}{!}{\includegraphics[angle=-90]{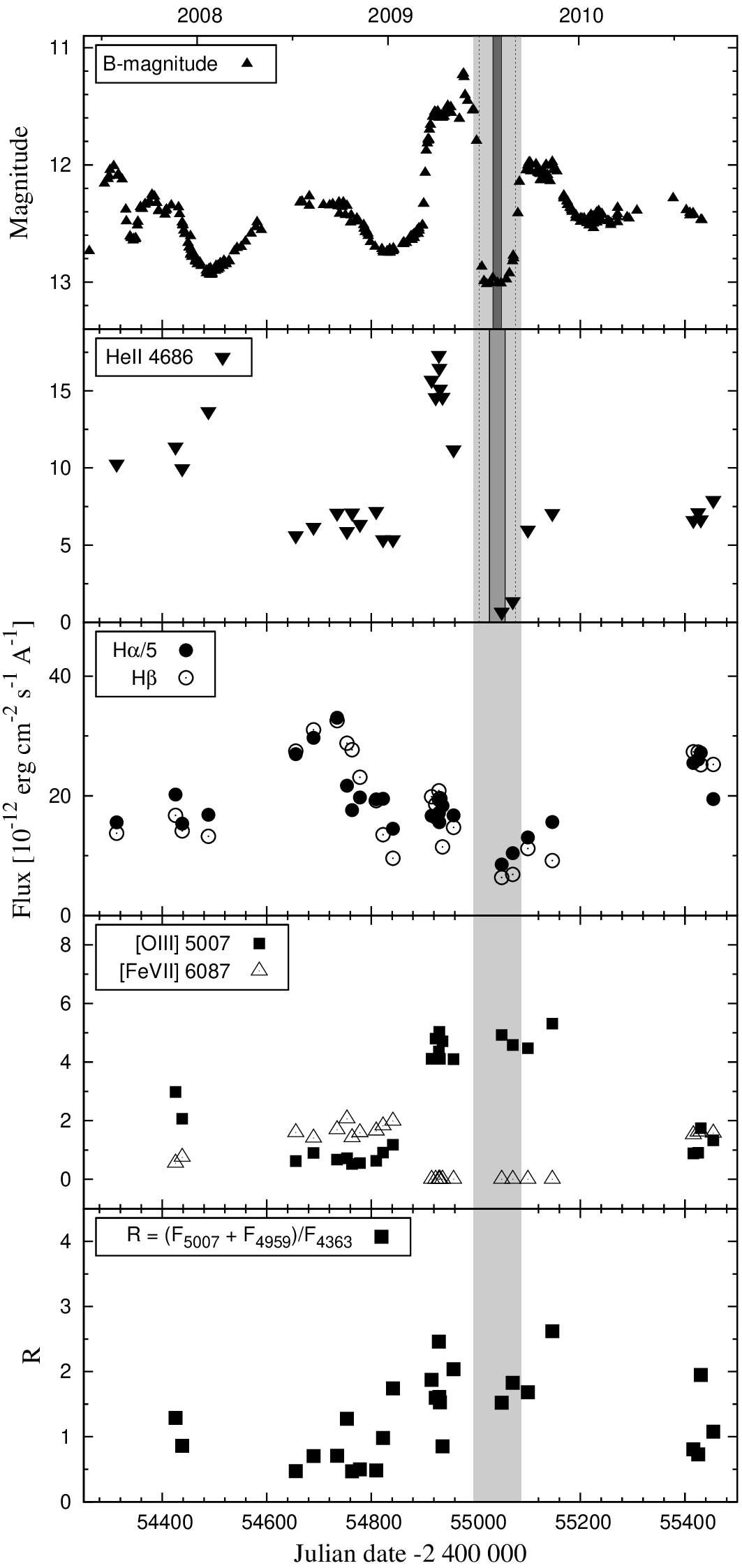}}
\caption[]{
Variation in the selected line fluxes during the 2007-10 active 
phase. Top panel shows the $B$-band LC to compare variation in 
the continuum and to visualize better the time-interval 
of the eclipse (the light shadow band). The width of the dark 
and mild shadow band corresponds to the linear size of the eclipsed 
object and the ${\rm He}^{++}$ zone, respectively. Dotted lines 
limit the radius of the cool giant. 
Fluxes are dereddened and their values are given in Table~5. 
          }
\end{center}
\end{figure}

{\sf \ion{He}{ii}\,4686\,\AA\ line:}
The \ion{He}{ii}\,4686\,\AA\ emission line nearly disappeared 
during the eclipse. This locates the region of its origin 
in the hot star vicinity and constrains its maximum linear 
size to the diameter of the eclipsing giant. The measured flux 
in the \ion{He}{ii}\,4686 line, $F_{\rm 4686}$, allows us to 
estimate the mean electron concentration in the He$^{++}$ zone. 
Its luminosity can be expressed as 
\begin{equation}
 4\pi d^2 F_{\rm 4686} = \alpha_{4686}(T_{\rm e}) n_{\rm e} 
                         n({\rm He}^{++}) h\nu_{4686} V_{4686}, 
\end{equation}
where $\alpha_{4686}(T_{\rm e})$ is the effective recombination 
coefficient for the given transition, 
$n_{\rm e}$ and $n({\rm He}^{++})$ is the mean concentration of 
electrons and  He$^{++}$ ions, respectively, and $V_{4686}$ is 
the volume of the He$^{++}$ zone. We also assume that all 
\ion{He}{} atoms are fully ionized within the zone, i.e. the 
abundance $A({\rm He}^{++}$) = $A(\ion{He}{})$, and thus 
$n({\rm He}^{++})$ = $A(\ion{He}{})n_{\rm p}$, because the 
concentration of hydrogen atoms is equal to that of protons, 
$n_{\rm p}$, in the ionized zone. Then the electron concentration 
\begin{equation}
 n_{\rm e} = (1 + 2 A({\rm He}^{++})) n_{\rm p}, 
\end{equation}
because each \ion{He}{} atom produces two electrons within the 
He$^{++}$ region. For $A({\rm He}^{++})$ = 0.1 the ratio 
$n_{\rm p}/n_{\rm e} \sim 0.83$, and Eq.~(2) can be 
expressed as 
\begin{equation}
 4\pi d^2 F_{\rm 4686} = \left(\frac{4}{3}\pi R_{\rm g}^3\epsilon\right)
  \alpha_{4686}(T_{\rm e})A(\ion{He}{}) n_{\rm e}^2 0.83 h\nu_{4686},
\end{equation}
where we approximated the $V_{4686}$ volume by a sphere with 
the radius of the eclipsing giant, $R_{\rm g}$, and $\epsilon$ 
is the filling factor. 
Our measured fluxes, 
 $F_{\rm 4686} \sim 1.56 \times 10^{-11}$ and 
               $\sim 7 \times 10^{-12}$\ecs, 
produced by the He$^{++}$ zone before and after the eclipse,
respectively (Fig.~4, Table~5), 
$\alpha_{4686}(30\,000\,{\rm K}) h\nu_{4686}$ =
$4.2\times 10^{-25}$\,\rec\ \citep[][]{h+s87}, 
$R_{\rm g} = 115$\ro\ (Sect.~3.1.2.) and $\epsilon = 1$, 
yield a mean $n_{\rm e} = 8.6\times 10^{9}$ and 
$5.7\times 10^{9}$\cmt\ during the 2009 brightening, 
before and after the eclipse, respectively. 
These quantities represent a minimum $n_{\rm e}$, because 
an optically thin case is assumed. 
The size of the He$^{++}$ zone can also be estimated from 
spectroscopic observations made close to the $T_3$ time 
(spectrum 2009/08/26) and just after the $T_4$ time (2009/09/24), 
when the \ion{He}{ii}\,4686 line also aroused from the eclipse 
(Fig.~4). This time interval ($\sim$30 days) corresponds to 
$R_{4686} \sim 50$\,\ro, which yields 
$n_{\rm e} \sim 2.0\times 10^{10}$\cmt\ (volume of the eclipsed 
object was not subtracted). 

Profile of the \ion{He}{ii}\,4686\,\AA\ line was not symmetrical 
with respect to the reference wavelength. The RV of its peak was 
placed mostly at the antiphase to the RV curve of the giant 
(Fig.~5), which supports the connection of the He$^{++}$ zone 
to the hot star surroundings. Its larger amplitude indicates 
an extension of the He$^{++}$ zone to farther distances from 
the hot star. However, the position of the main core of the 
profile, measured around the half of its maximum, was shifted 
blueward relative to the systemic velocity 
\citep[see also Fig.~5 of][]{mk92}. 
The violet shift of the \ion{He}{ii}\,4686\,\AA\ emission line 
was measured also in the spectra of other symbiotics. For example, 
\cite{tamura+03} measured it for Z~And and \cite{mu+09b} 
for AG~Dra. \cite{sk+06} interpreted this effect by the presence 
of a disk around the hot star, which blocks a part of the radiation 
from the wind at its backside to the direction of the observer. 
Finally, the profile seems to be affected by a variable absorption 
from its violet side, similarly as for the hydrogen lines (Fig.~3), 
which makes it difficult to interpret the \ion{He}{ii}\,4686\,\AA\
line profile more accurately. 
%
%
%
\addtocounter{table}{+1}
\begin{table}
\caption[]{Radial velocities in \kms\ of the main emission peak 
           of highly ionized elements. The systemic velocity 
           was subtracted.} 
\begin{center}
\begin{tabular}{cccc}
\hline
\hline
Date & \ion{He}{ii}\,$\lambda$4686 
     & [\ion{Fe}{vii}]\,$\lambda$6087
     & [\ion{O}{iii}]\,$\lambda$5007 \\
\hline
2007/07/31 &  10.5 &   --   &  --    \\
2008/01/23 &   2.0 &   --   &  --    \\
2008/10/13 &   6.9 &  10.3  &  27.1  \\
2008/12/21 &  20.3 &  24.8  &  33.4  \\
2009/01/09 &  22.7 &  25.8  &  31.7  \\
2009/04/14 &  23.8 &   --   & -16.2  \\
2009/11/10 & -18.0 &   --   & -14.1  \\
2010/08/21 &  -0.6 &  -13.4 &  29.4  \\
\hline
\end{tabular}
\end{center}
\end{table}    
%

{\sf Forbidden lines:}
In contrast to the \ion{H}{} and \ion{He}{} lines, the nebular 
$N_1$ and $N_2$ ([\ion{O}{iii}]\,5007 and 4959\,\AA) lines 
were not subject to the eclipse. Fluxes of the stronger 
5007\,\AA\ line were as low as $\approx 1\times10^{-12}$\ecs\ before 
2009 March, when they increased to $\sim 4.5\times 10^{-12}$\ecs, 
following the increase in the optical continuum. However, during 
the eclipse the line flux persisted at the same level. 
In 2010 it returned to the pre-brightening values (Fig.~4). 
The ratio $R = F(N_1 + N_2)/F_{4363}$, which is a well-known probe 
of $n_{\rm e}$ and $T_{\rm e}$ in planetary nebulae, was extremely 
low. The observed values, 0.5 -- 2, imply a high-density 
[\ion{O}{iii}] nebula with 
$n_{\rm e}([\ion{O}{iii}]) \sim 10^{7} - 10^{8}$\,\cmt\ and 
$T_{\rm e}([\ion{O}{iii}]) < 20\,000$\,K 
\citep[e.g. Fig.~15.1 in][]{gurzadyan}. A detailed application 
of the method to AX~Per can be found in \cite{sk+01}, 
who derived the upper limit of $n_{\rm e}([\ion{O}{iii}])$ 
as $7\times 10^{7}$\,\cmt\ for $R = 1.7 - 4.3$. 
It is interesting that during the 1994 eclipse fluxes 
in the $N_1$ and $N_2$ lines faded by a factor of $\le $2 with 
respect to their out-of-eclipse values (see their Fig.~6 and 
Table~4), in contrast to the evolution during the 2009 eclipse. 
We discuss this effect in Sect.~4.1.1. 

The highly ionized [\ion{Fe}{vii}]\,6087\,\AA\ line was present 
in the spectrum only during a lower activity level. From the 
optical brightening in 2009 March to our observation on 2009 
November 10th, it entirely disappeared from the spectrum 
(Table~5). 
Its profile was asymmetrical with respect to the reference 
wavelength, being broadened more to the violet side of the line 
(Fig.~3). As a result, the RV of the intermediate point at half 
of the line intensity was negative relative to the systemic 
velocity (--18.1, --9.0, --7.3 and --13.4\kms\ as on 2008/10/13, 
2008/12/21, 2009/01/09 and 2010/08/21, respectively). However, 
the peak of the emission line followed the RV curve as derived by 
\cite{iijima88}; (see Fig.~5 here), which implies that a part of 
the [\ion{Fe}{vii}] zone has to be connected with the hot star, 
but it is located farther from it, behind the binary. 
The [\ion{Fe}{vii}] transitions are excited 
by collisions with thermal electrons \citep[][]{no70}. 
The enhanced ionized wind during the brightening produces a 
larger amount of free electrons, while the 
[\ion{Fe}{vii}]\,6087\,\AA\ emission line disappears. 
This anti-correlation can be understood if the [\ion{Fe}{vii}] 
region is placed instead at the orbital plane, where the material 
is in the ionization shadow during the phase of a warm disk 
creation (see Sect.~4). During quiescence, when the disk 
is not present, the [\ion{Fe}{vii}] line can be created within 
the collisional zone of winds from the binary components 
\citep[][]{wall+84}. According to its shaping in the binary 
\citep[see Fig.~5 of][]{bis+06}, a fraction of the [\ion{Fe}{vii}] 
emission could arise behind the binary from the hot star side, 
and thus satisfy the RV of its orbital motion with 
a larger amplitude (see Fig.~5). 
However, more high-resolution observations of the
[\ion{Fe}{vii}]\,6087\,\AA\ line for eclipsing symbiotic 
binaries are needed to understand its origin in symbiotic 
systems better. 
%
%
%
\begin{figure}
\centering
\begin{center}
\resizebox{\hsize}{!}{\includegraphics[angle=-90]{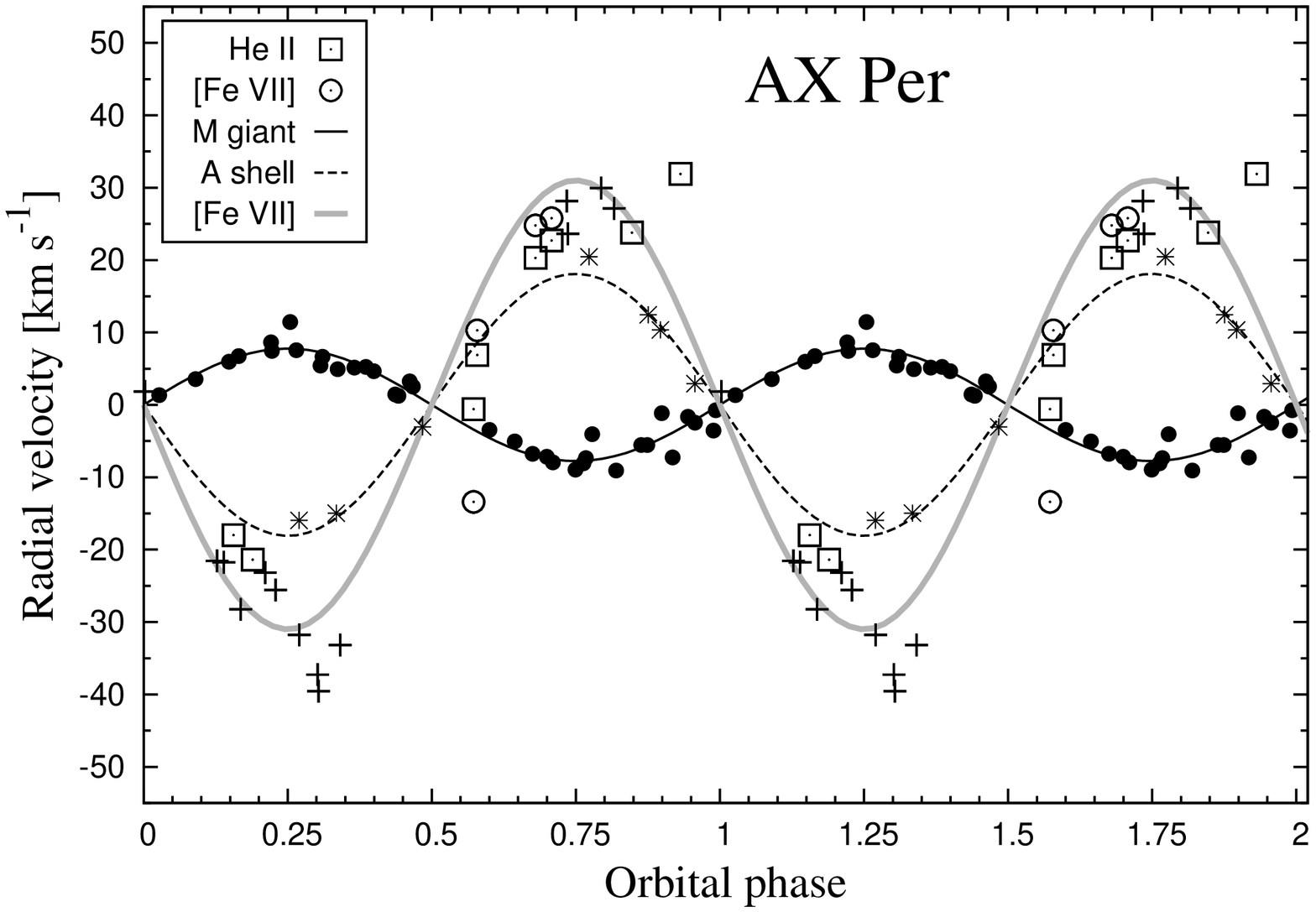}}
\caption[]{
Radial velocities of main emission peaks measured for highly 
ionized elements (open symbols, Table~6) as a function of 
the photometric phase (1). Compared are radial velocity curves 
for the M-type \citep[][ solid line, filled circles]{fekel+00} 
and A-type absorption lines \citep[][ dashed line, asterisks]{mk92}, 
and emission peaks of [\ion{Fe}{vii}]\,$\lambda$6087 line 
\citep[][ grey line, crosses]{iijima88}. Systemic velocity of 
-117.44\kms\ was subtracted. 
          }
\end{center}
\end{figure}

\subsubsection{Mass loss through the hot star wind}

The hot components in symbiotic binaries can lose their mass in 
the form of wind \citep[e.g.][]{vogel93,nsv95,eriksson,kenny05}, 
which enhances considerably during active phases 
\citep[][]{sk06}. To estimate its rate, we used the \ha\ method 
as proposed by \cite{sk06}, which assumes that the broad \ha\ 
wings originate in the ionized wind from the hot star. 
The principle of the method is a comparison of the observed 
and synthetic profile of the broad wings. 

The model assumes a spherically symmetric wind originating in 
the central star, which is covered by an optically thick disk 
in the direction of the observer. The disk is characterized 
with the radius $R_{\rm D}$ and height $H_{\rm D}$, and it is 
seen edge-on in the model (see Fig.~9). 
Density, and thus the emissivity of the wind at a given 
distance $r$ from the hot star centre, is determined by 
the mass-loss rate and the velocity of the wind. The velocity 
distribution is approximated with \citep[][]{cak} 
%
\begin{equation}
 v(r) = v_{\infty}(1 - R_{\rm w}/r)^{\beta}, 
\end{equation}
%
where the origin of the wind, $R_{\rm w}$, and $\beta$ are 
model parameters, while the terminal velocity, $v_{\infty}$, 
is given by the extension of the wings. 
During the quiescent phase, parameters $R_{\rm D}$ and $H_{\rm D}$ 
can be estimated from the effective radius of the hot star, 
$R_{\rm h}^{\rm eff} \sim 0.1$\,\ro\ \citep[Table~3 of][]{sk05}. 
Assuming the ratio $H_{\rm D}/R_{\rm D} = 0.1$ yields
$R_{\rm D} = 0.32$\,\ro\ and $H_{\rm D} = 0.032$\,\ro. 
During the investigated active phase, we estimated the disk 
radius from the first two contact times of the 2009 eclipse 
(Sect.~3.1.2.) as $R_{\rm D} \sim 20$\,\ro. Assuming a disk 
with $H_{\rm D}/R_{\rm D} = 0.3$, yields $H_{\rm D}=6$\,\ro. 
We recall here that the results do not critically depend 
on these parameters. The mass-loss rate is mainly determined 
by the luminosity of the broad wings and their terminal 
velocity \citep[see Eq.~(14) of][]{sk06}. 
A comparison between the modeled and observed profiles is 
shown in Fig.~6. The synthetic profiles match the observed 
wings well for $|RV| \ga 200-250$\kms. There 
is a significant difference between the extension and 
luminosity of the \ha\ wings observed during the quiescent and 
active phase (Table~7, Fig.~6). Broad wings from the 1998 
quiescence corresponded to the mass-loss rate of 
$\la 1\times 10^{-7}$\myr, while during the activity 
it increased to $\approx 2-3\times 10^{-6}$\myr. 

In our model the mass-loss rate determines the particle density 
at a given distance $r$ from the origin of the wind according 
to the continuity equation as 
\begin{equation}
  n_{\rm w}(r) = \dot M_{\rm w}/4 \pi r^2 \mu m_{\rm H} v(r),
\end{equation}
where the wind velocity, $v(r)$, is given by Eq.~(5), $\mu$ is 
the mean molecular weight and $m_{\rm H}$ is the mass of 
the hydrogen atom. Within the He$^{++}$ zone, i.e. from the 
hot star to $r_{\rm HeII} \approx 50$\ro\ (Sect.~3.2.1), 
the particle densities $n_{\rm w}(r)$ excellently agree 
with those derived independently of the measured $F_{\rm 4686}$ 
fluxes. For example, $\dot M_{\rm w} = 3\times 10^{-6}$\myr\ 
and $v_{\infty} = 2000$\kms, yield 
$n_{\rm w}(20\,R_{\sun}) \sim 2.8 \times 10^{10}$ 
(i.e. $n_{\rm e} = n_{\rm w}/0.83 = 3.3\times 10^{10}$\cmt), 
and $n_{\rm w}(50\,R_{\sun}) \sim 3.2 \times 10^{9}$\cmt\ 
($n_{\rm e} = 3.9\times 10^{9}$\cmt). 
This suggests that particles of the hot star wind give rise 
to the observed emission from the He$^{++}$ zone. 
%
%
\begin{figure}
\centering
\begin{center}
\resizebox{8cm}{!}{\includegraphics[angle=-90]{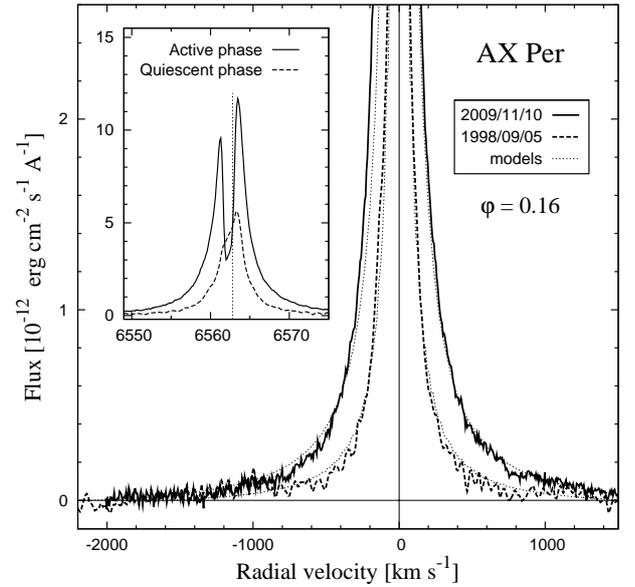}}
\caption[]{
Example of the broad \ha\ wings from the 2007-10 active phase 
(solid line), compared with those observed during quiescent 
phase (dashed line). Both spectra were taken at the same orbital 
phase, $\varphi = 0.16$. Local continua were subtracted from 
the spectra. The spectra were dereddened. 
The models (dotted lines) indicate a significant enhancement of 
the mass-loss rate during activity with respect to quiescence 
(Table~7, Sect.~3.2.2). 
          }
\end{center}
\end{figure}
%
%
%
\begin{table*}
\caption[]{
Parameters of synthetic models of the broad \ha\ wings 
($R_{\rm w}$, $\beta$, $v_{\infty}$), their luminosities 
observed for $|RV| \ge 200$\kms, $L_{\alpha}$(200), 
and corresponding mass-loss rates, $\dot M_{\rm w}$. 
          }
\begin{center}
\begin{tabular}{cccccc}
\hline
\hline
  Date           &  
$R_{\rm w}$      &  
$\beta$          &  
$v_{\infty}$     &  
$L_{\alpha}$(200)&  
log($\dot M_{\rm w}$)    \\
dd/mm/yyyy       &  
[$R_{\sun}$]     &  
                 &  
[km\,s$^{-1}$]   &  
[$L_{\sun}$]     &  
[$M_{\sun}\,\rm yr^{-1}$] \\
\hline\\[-3mm]
\multicolumn{6}{c}{Quiescent phase: 
                   $R_D = 0.32$\ro, $H_D = 0.032$\ro}\\[1mm]
1998/01/09 & 0.027 & 1.75 & 1500 & 0.90 & -7.04 \\
1998/09/05 & 0.030 & 1.75 & 1600 & 0.88 & -7.02 \\
\multicolumn{6}{c}{Active phase: 
                   $R_D = 20$\ro, $H_D = 6$\ro}\\[1mm]
2007/07/31 & 5.20 & 1.75 & 2500 & 2.97 & -5.52 \\
2008/01/23 & 5.27 & 1.74 & 2500 & 2.13 & -5.59 \\
2008/10/13 & 5.40 & 1.75 & 1500 & 0.83$^\dagger$ & -5.84 \\
2008/12/21 & 5.45 & 1.72 & 2000 & 1.60 & -5.70 \\
2009/01/09 & 5.40 & 1.75 & 2000 & 1.40 & -5.72 \\
2009/04/14 & 5.25 & 1.73 & 2500 & 3.17 & -5.49 \\
2009/11/10 & 5.20 & 1.75 & 2100 & 1.42$^\dagger$ & -5.67 \\
2010/08/21 & 0.94$^\ddagger$ & 1.80 & 2000 & 2.25$^\dagger$ & -5.96 \\
\hline
\end{tabular}
\end{center} 
 $^\dagger$ $|RV| = 250$\kms, 
 $^\ddagger$ $R_D = 3.5$\ro, $H_D = 1.05$\ro
\end{table*}

\subsubsection{Modeling the continuum spectrum}

The observed continuum spectrum of symbiotic stars, $F(\lambda)$, 
can formally be expressed as a superposition of its three 
basic components (see Sect.~1), i.e. 
%
\begin{equation}
  F(\lambda) =  F_{\rm g}(\lambda) + F_{\rm n}(\lambda) +
                F_{\rm h}(\lambda), 
\end{equation}
%
where $F_{\rm g}(\lambda)$, $F_{\rm n}(\lambda)$ and 
$F_{\rm h}(\lambda)$ represent radiative contributions from 
the cool giant, nebula and a hot stellar source, respectively. 
Basic features of these components of radiation are well 
recognizable in our low-resolution spectra exposed from short
wavelengths, $\lambda \ga 3300$\,\AA. Most distinctive are 
molecular passbands arising in the red giant atmosphere, while 
a pronounced Balmer jump in emission directly indicates 
a strong nebular continuum. The source of ionizing 
photons (i.e. the hot star) is indicated indirectly by 
the presence of strong recombination lines superposed on 
the continuum. Its direct indication is possible only within 
the super-soft X-ray and far-UV; in the optical its contribution 
is negligible \citep[e.g. Fig.~2 of][]{sk+09}. 
Another significantly cooler stellar source of radiation 
develops during active phases. It is connected with the hot 
star, and its evidence is given directly by eclipses in the 
optical LCs \cite[e.g.][]{bel79,sk91} and/or by model SED at 
any other orbital phase \citep[][]{sk05}. 
Hereafter we call it a warm stellar component (WSC). 

The aim of this section is to disentangle the composite spectrum, 
observed at different stages of AX~Per activity, into its individual 
components of radiation, i.e to determine their physical parameters. 
For the $F_{\rm g}(\lambda)$ component we used photospheric 
synthetic spectra of M-giant stars published by \cite{fluks}. 
These models were calculated in the spectral range 350--900\,nm 
for 11 spectral types (ST), from M0 to M10. In our modeling 
we determine the ST and its subclass by a linear interpolation 
between the neighboring best-fitting STs. Because the radiation from 
the giant can vary, the scaling factor of the synthetic spectra 
represents another parameter in the model SED. 

We approximated the spectrum produced by the nebula by 
\begin{equation}
  F_{\rm n}(\lambda) = k_{\rm n}\times 
                       \varepsilon_{\lambda}(T_{\rm e}),
\label{eqn:neb}
\end{equation}
where the scaling factor $k_{\rm n}$ [cm$^{-5}$] represents the 
emission measure (\textsl{EM}) of the nebula scaled with the 
distance $d$, i.e., $\textsl{EM}=4\pi d^2\times k_{\rm n}$. 
$\varepsilon_{\lambda}(T_{\rm e})$ is the volume emission 
coefficient [${\rm erg\,cm^3\,s^{-1}\,\AA^{-1}}$], which depends 
on the electron temperature, $T_{\rm e}$, and is a function of 
the wavelength \citep[e.g.][]{b+m70}. By this approach we can 
register only those photons produced by the optically thin part 
of the nebular medium, and thus estimate the lower limit 
of $EM$. For the sake of simplicity, we calculated 
the $\varepsilon_{\lambda}(T_{\rm e})$ coefficient 
for the hydrogen plasma only, including contributions from 
recombinations and thermal bremsstrahlung. Some additional 
arguments for this approximation are given in \cite{sk+09}. 
In addition, we assumed that $T_{\rm e}$ and thus 
$\varepsilon_{\lambda}(T_{\rm e})$ are constant throughout
the nebula. 
%
%
\begin{figure}
\centering
\begin{center}
\resizebox{\hsize}{!}{\includegraphics[angle=-90]{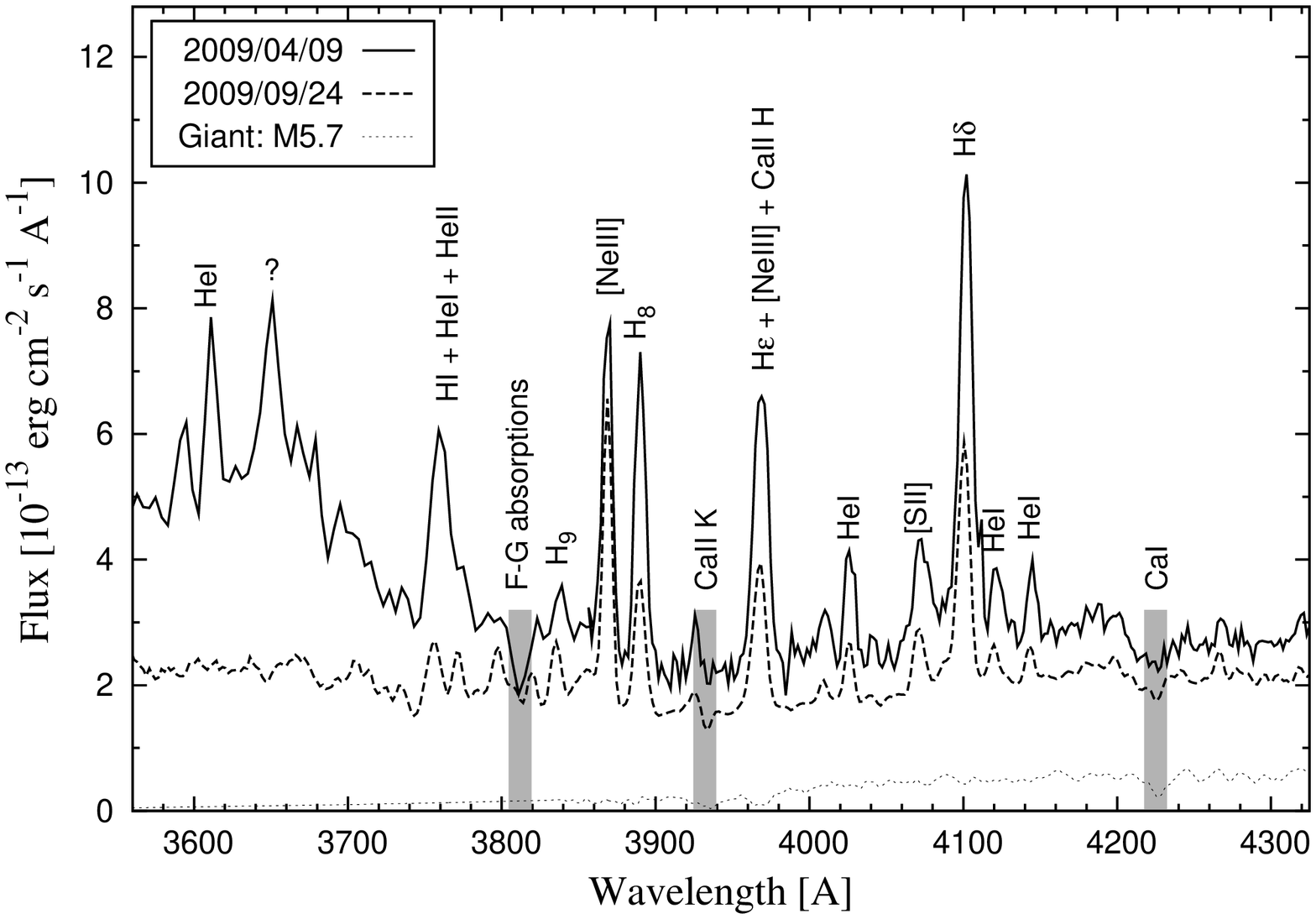}}
\caption[]{
Blue part of the dereddened spectra with well-developed 
warm pseudophotosfere as suggested by the model SED (Fig.~8). 
Features typical for the spectrum of F-G giants are marked 
by gray bands. 
          }
\end{center}
\end{figure}
%

The presence of F-G type features in the spectra before and 
after the 2009 eclipse suggests that the temperature of the stellar 
source is as low as $\approx 5000 - 7000$\,K. As the most obvious 
F-G features we identified the broad \ion{Ca}{i}\,$\lambda$4227 and 
\ion{Ca}{ii}\,K\,$\lambda$3934 line (Fig.~7). Also the absorption 
at $\lambda$3810, pronounced mainly in the 2009/04/09 spectrum, 
is clearly evident in the F and G spectral templates showed on 
Figs.~218 to 222 of \cite{m+z02}. 
Therefore, we compared the continuum of the hot stellar source 
in Eq.~(7) to synthetic spectra, calculated for 
5\,000 -- 10\,000\,K using the Kurucz's codes \citep[][]{munari+05}. 
So, the third term of Eq.~(7) can be written as 
\begin{equation}
 F_{\rm h}(\lambda) = (\theta_{\rm warm}^{\rm eff})^2\, 
                      \mathcal{F}_{\lambda}(T_{\rm eff}),
\end{equation}
where $\mathcal{F}_{\lambda}(T_{\rm eff})$ is the WSC 
of radiation. 
Its corresponding effective temperature, $T_{\rm eff}$, 
and the scaling factor, 
$\theta_{\rm warm}^{\rm eff} = R_{\rm warm}^{\rm eff}/d$, 
represent free parameters in modeling the SED. 
Other atmospheric parameters of synthetic spectra were fixed 
($\log(g) = 2.5, [M/H] = 0, [\alpha /{\rm Fe}] = 0, 
v_{\rm rot} = 20$\,\kms), and their resolution was accommodated 
to that of our low-resolution spectra. 
%
%
%
\begin{figure*}
\centering
\begin{center}
%
\resizebox{\hsize}{!}{\includegraphics[angle=-90]{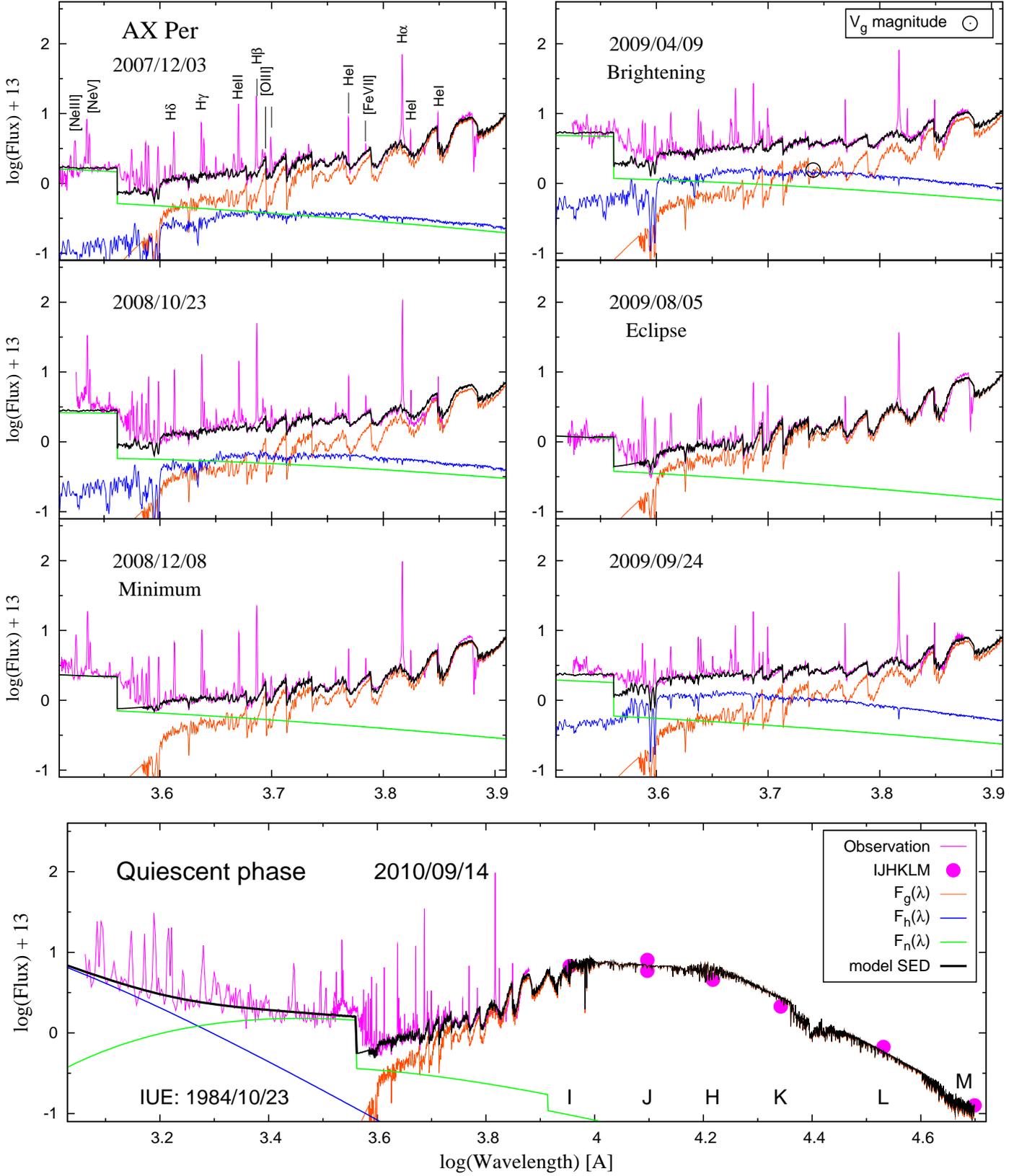}}
\caption[]{
Our low-resolution dereddened spectra (violet lines) and 
their models (heavy black lines) at selected dates during the 
2007-10 active phase of AX~Per. 
The model SEDs and their components of radiation here 
represent a graphic form of Eq.~(7) with the same denotation in 
keys (in the bottom panel). Fluxes are in units \ecsa. 
Corresponding parameters are collected in Table~8. 
Modeling is described in Sect.~3.2.3. 
          }
\end{center}
\end{figure*}
%

After defining the individual components of radiation in Eq.~(7), 
we prepared a grid of models for reasonable ranges of the 
fitting parameters (ST, its subclass and scaling for the giant; 
$T_{\rm e}$ and $k_{\rm n}$ for the nebular continuum; 
$T_{\rm eff}$ and $\theta_{\rm warm}^{\rm eff}$ for the WSC), 
and selected that which corresponded to a minimum of the function 
%
%
\begin{equation}
 \chi^{2} = \sum_{\rm i=1}^{\rm N}
    \left[\frac{(F^{\rm obs}(\lambda_{\rm i}) -
            F(\lambda_{\rm i})}
            {\Delta F^{\rm obs}(\lambda_{\rm i})}\right]^{2},
\label{eq:chi2}
\end{equation}
%
where $F^{\rm obs}(\lambda_{\rm i})$ are fluxes of the observed 
continuum, $N$ is their number (1500 -- 1800), 
$\Delta F^{\rm obs}(\lambda_{\rm i})$ are their errors, 
and $F(\lambda_{\rm i})$ are theoretical fluxes given by 
Eq.~(7). Input flux-points 
were selected from the observed spectra by omitting emission 
lines and a spectral region from 3645 to $\sim$3740\,\AA, where 
hydrogen lines of the high members of Balmer series were 
blended, which did not allow us to identify the true continuum. 
For the flux uncertainties we adopted typical values of 10\%. 

Our resulting models of the AX~Per composite continuum along 
its 2007-10 active phase are shown in Fig.~8 and the corresponding 
parameters are given in Table~8. The presence of the WSC 
in the 3200--7500\,\AA\ spectrum is constrained 
by a low Balmer jump and high values of the observed fluxes 
in the blue part of the spectrum. To fit the small Balmer 
discontinuity and to fill in the large difference between 
the contribution from the giant and the observed spectrum 
for $\lambda \la 5000$\,\AA, an additional radiative component 
with a fairly flat energy distribution throughout the optical 
was required. This component satisfies the radiation from 
a stellar source with temperature of 5000-6000\,K well, 
and can accordingly be associated with the WSC. 
The WSC and the nebular emission were strongest during 
the 2009 brightening (Table~8). 
The following eclipse interrupted the brightening by removing 
the WSC entirely and the nebular component partially from 
the spectrum (\textsl{EM} decreased by a factor of $\ga$3). 
Immediately after the eclipse, the WSC arose again in 
the spectrum together with a stronger nebular emission 
(Fig.~8 and Table~8). We were able to fit our last spectra from 
2010 August and September without the WSC, which signaled 
the return of AX~Per to quiescent phase. 
For comparison, we inserted the last spectrum into the UV-IR SED 
from quiescence \citep[adapted according to Fig.~21 of][]{sk05} 
to demonstrate the negligible contribution of the hot 
star in the optical (Fig.~8, bottom). 

For comparison, we also derived $T_{\rm e}$, $EM$ and the 
intrinsic $V$-magnitude of the giant, $V_{\rm g}$, from the 
observed $UBV$ magnitudes by using the method of \cite{c+s10}. 
For the 2009 eclipse we used $B = 13.01$ and $V = 11.59$, 
taken simultaneously with the low-resolution spectrum from 
2009/08/05. The $U$ magnitude was interpolated to 12.93 at 
this date. After simultaneous corrections for emission lines 
(Table~4), we determined $T_{\rm e} \sim 37000$\,K, 
$EM = 2.7\pm 0.2 \times 10^{59}\,(d/1.73\,\kpc)^2$\cmt\ and 
$V_{\rm g} = 10.98\pm 0.02$. Uncertainties of the resulting 
parameters reflect errors of photometric measurements. 
The $V_{\rm g}$ flux-point matches the contribution from 
the giant well (Fig.~8), the $EM$ is also well comparable with 
that derived from the model SED (Table~8), but $T_{\rm e}$ 
suffers from a large uncertainty caused by a weak dependence 
of the nebular continuum profile on $T_{\rm e}$ within 
the $UBV$ region only. 
%
%
%
\begin{table*}
\caption[]{Parameters of the SED-fitting analysis
           (see Sect.~3.2.3, Fig.~8).
          }
\begin{center}
\begin{tabular}{cc|ccc|ccc|cc|c}
\hline
\hline
  Date                                   &
 Stage/B-mag                             &
\multicolumn{3}{c|}{Giant}               &
\multicolumn{3}{c|}{Warm stellar object} &
\multicolumn{2}{c|}{Nebula}              &
$\chi^2_{\rm red}$                       \\
yyyy/mm/dd                        &
                                  &
ST                                &
$T_{\rm eff}$/K$^\dagger$         &
$L_{\rm g}/L_{\sun} ^\star$      &
$T_{\rm eff}$/K                   &
$R_{\rm warm}^{\rm eff}/R_{\sun}$ &
$L_{\rm warm}/L_{\sun}$           &
$T_{\rm e}$/K                     &
$EM/10^{59}$\cmt                  &
                                  \\
%
%
\hline
2007/12/03& A / 12.5 & M5.8 & 3316& 1430 & 5250 & 6.2  & 26 & 34\,000 & 3.5 & 0.75 \\
2008/10/23& A / 12.4 & M5.8 & 3316& 1430 & 5250 & 8.2  & 45 & 23\,000 & 3.8 & 1.17 \\
2008/12/08& A / 12.7 & M5.8 & 3316& 1430 &  --  &  --  & -- & 32\,000 & 4.8 & 0.87 \\
2009/04/09& A / 11.6 & M5.7 & 3326& 1450 & 5500 &  11  &  98& 25\,000 & 8.2 & 0.80 \\
2009/08/05& A / 13.0$^\ddagger$ &M5.8 & 3316& 1430 & --  &  --  & -- & 33\,000 & 2.6 & 0.78 \\
2009/08/26& A / 12.8$^\ddagger$ &M5.6 & 3336& 1470 & --  &  --  & -- & 28\,000 & 1.7 & 1.06 \\
2009/09/24& A / 12.0 & M5.6 & 3336& 1470 & 6250 &  7.0 & 68 & 32\,000 & 4.0 & 0.55 \\
2010/08/16& Q / 12.4 & M5.3 & 3366& 1520 & --  &  --  & -- & 27\,000 & 4.0 & 1.30 \\
2010/09/14& Q / 12.5 & M5.3 & 3366& 1520 & --  &  --  & -- & 27\,000 & 3.5 & 1.67 \\
\hline
\end{tabular}
\end{center} 
 $^\dagger$ according to the \cite{fluks} calibration,~
 $^\star$ for $R_{\rm g} = 115$\ro,~
 $^\ddagger$ eclipse 
\end{table*}
\section{Discussion}

From 2007 on, AX~Per entered a new active phase (Sect.~3.1.1). 
This was also supported by a correlation between the visual 
brightness, $EM$ and the WSC luminosity (see Table~8), as 
observed for the hot components during active phases in other 
symbiotic stars \citep[see Table~4 of][]{sk05}. 
The 2007-10 active phase was connected with a significant change 
in the ionization structure in the binary during the transition 
from the preceding quiescent phase. 

(i) 
In the continuum, this was indicated by a distinctive 
change in the shape of the minimum observed at/around 
the inferior conjunction of the giant. 
During quiescence, the broad minima in the LC (e.g. E = 4 to 9 
in Fig.~1) are caused by the orbital motion of extended 
ionized wind from the giant, which is partially optically thick 
\citep[][]{sk01}. During the active phase, the broad minima were 
replaced by narrow ones (eclipses), which suggests that 
a large part of the optical light produced by the hot 
component was concentrated at/around the hot star. 

(ii)
In the line spectrum, we indicated a significant increase in the 
mass-loss rate via the hot star wind with respect to the quiescent 
phase (Sect.~3.2.2, Fig.~6, Table~7). For example, during the 2009 
brightening, we measured the highest luminosity and extension of 
the \ha\ wings, which corresponded to the mass-loss rate of 
$\sim 3\times 10^{-6}$\myr. 

\subsection{Evolution of the ionized zone}

Throughout the whole 2007-10 period we observed a strong nebular 
component of radiation in the spectrum (Fig.~8, Table~8). 
A geometrical change of the main source of the nebular 
radiation was indicated by the change in the position and profile 
of eclipses during the 2007 and 2009 inferior conjunction of the 
giant. The 2007 minimum was shifted from the conjunction by 
$\sim -0.025 P_{\rm orb}$ (Fig.~2, right). 
This implies that an enhanced emitting region had to be located 
in front of the hot star, preceding its orbital motion. 
As discussed by \cite{bis+06}, this higher density 
region can develop about 100-150 days after an increase 
of the hot star wind as a result of its collision with the 
giant's wind and the orbital motion (see their Figs.~4 and 5). 
This region will be ionized by the neighboring hot star, 
and thus produce a surplus of nebular emission. As a result, 
the light minimum will precede the inferior conjunction of 
the giant. However, after the following orbital cycle, the 
minimum was located exactly at the giant's conjunction 
(Fig.~2). Changes in the position of the light minima 
during different stages of activity were indicated also for 
other eclipsing symbiotic binaries \citep[see][]{sk98}. 

During the 2009.6 eclipse the WSC of radiation disappeared 
entirely, and the nebular component faded by a factor of 
$\sim$3 with respect to the out-of-eclipse values 
(Table~8, Fig.~8). This implies that a strong source of the 
nebular continuum had to be located in the vicinity of the hot 
star, within the region eclipsed by the giant. 
According to the modeling of the broad \ha\ wings (Sect.~3.2.2), 
the eclipsed emission region can be ascribed to the ionized 
wind from the hot star. 

\subsubsection{Nebular lines} 

Spectroscopic observations made after the major 1989-91 active 
phase at/around the 1994 eclipse showed that the nebular 
[\ion{O}{iii}] lines were subject to eclipse. Their fluxes faded 
by a factor of $\sim$2 with respect to values measured before 
and after the eclipse \citep[see Table~4 of][]{sk+01}. In contrast, 
from 2009 March, nebular [\ion{O}{iii}] line fluxes increased 
by a factor of $\sim$4.5 and kept their values at a constant level 
to our last 2009 observation (2009/11/10), i.e. also throughout 
the eclipse (Fig.~4). This behavior implies that the hot star 
wind causes a fraction of the [\ion{O}{iii}] region. 
Its location is therefore a function of the mass-loss rate, 
which determines particle densities at a given distance from 
the wind source. 
Higher $\dot M_{\rm w}$ places the [\ion{O}{iii}] zone to longer 
distances from the hot star and vice versa. During the active 
phase, $\dot M_{\rm w} \sim 2\times 10^{-6}$\,\myr, 
$v_{\infty} = 2000$\,\kms\ and critical densities for creation 
of nebular lines, $n_{\rm e} \ga 10^7$\,\cmt, 
place the [\ion{O}{iii}] zone at distances $r > 210$\,\ro\ from 
the hot star, i.e. far beyond the eclipsed region. However, 
during quiescence, $\dot M_{\rm w} \sim 1\times 10^{-7}$\,\myr\ 
and $v_{\infty} = 1600$\,\kms\ (Table~7) correspond to the 
critical radius of $\sim$50\,\ro, which thus can be partially 
eclipsed by the giant with $R_{\rm g} \sim 115$\,\ro. 
Finally, the [\ion{O}{iii}] line profiles consist of more 
components (Fig.~3), which suggests that the hot star wind 
at farther distances from its origin was structured and not 
simply uniform in direction and density. 
%
%
\begin{figure*}
\centering
\begin{center}
\resizebox{12cm}{!}{\includegraphics[angle=-90]{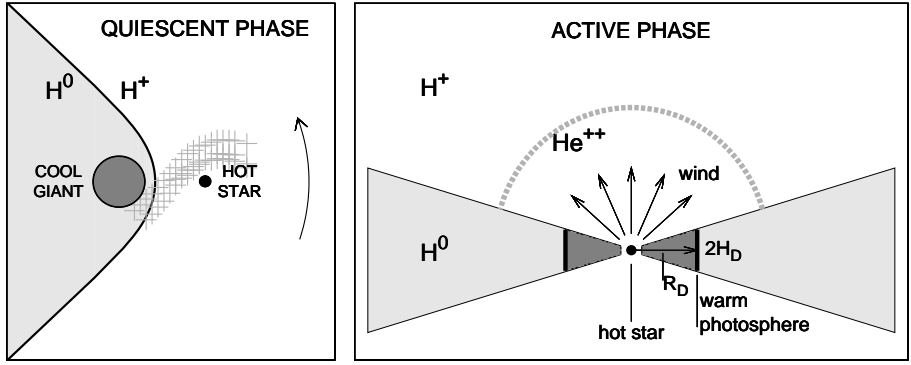}}
\caption[]{
Sketch of the ionization structure of AX~Per during the quiescent 
phase (left, pole-on view) and around the hot star during the 
active phase (right, edge-on view). 
During quiescence, the solid thick line represents the 
H$^{0}$/H$^{+}$ boundary in the steady-state approximation. 
Including the wind from the hot star and rotation of the binary 
(denoted by the arrow), a denser region evolves at the preceding 
front of the hot star, which is further twisted to the front of 
the giant against its motion 
\citep[crosses, adapted from Fig.~5 of][]{bis+06}. 
The right panel shows a sketch of the ionization structure of 
the hot component during activity. 
The neutral zone has the form of a flared disk, whose optically 
thick rim represents the warm pseudophotosphere. The enhanced 
wind from the hot star is ionized by its radiation. The He$^{++}$ 
boundary is denoted by the gray dashed line (see Sect.~4). 
          }
\end{center}
\end{figure*}

\subsection{The neutral zone}

We indicated the presence of a WSC of radiation in the composite 
spectrum photometrically by the 2009 eclipse profile, and 
spectroscopically by separating the observed spectrum into 
its individual components. Its source is characterized with 
effective temperatures of $\sim$5500 -- 6500\,K and effective 
radii of $\sim$6 -- 11\ro, which correspond to luminosities 
between $\sim$30 and 100\lo\ (Table~8). However, its shape 
cannot be spherical: 
(i) 
If this were a sphere, its radiation would not be capable 
of causing the observed nebular emission. 
On the other hand, the presence of the strong nebular component 
in the spectrum constrains the presence of a hot ionizing source 
in the system. 
(ii) 
If the radius of the eclipsed object, $R_{\rm e} = 27$\,\ro\ 
(Sect.~3.1.2), were that of a sphere, its luminosity would be 
a factor of $\sim$15 higher than that we observed at the 2009 
bright stage. These arguments suggest that the source of 
the WSC, which is subject to the eclipse, has a form of a disk 
with the radius $R_{\rm D} = R_{\rm e}$. Assuming that its 
outer rim represents the warm pseudophotosphere 
\citep[Sect.~5.3.5. of][]{sk05}, then its luminosity 
\begin{equation}
  2\pi R_{\rm D} 2 H_{\rm D} \sigma T_{\rm eff}^4 = 
  4\pi (R_{\rm warm}^{\rm eff})^2\sigma T_{\rm eff}^4 . 
\end{equation}
For $H_{\rm D}/R_{\rm D} \equiv 0.3$ (Sect.~3.2.2) and 
$R_{\rm warm}^{\rm eff} = 11$\ro\ (Table~8) we get 
$R_{\rm D} = 20$\,\ro, 
which agrees well with $R_{\rm e} = 22.3\pm 2$\,\ro, 
obtained from the first two contact times of the 2009 eclipse. 
Therefore we conclude that the warm stellar object, which developed 
during the active phase, has the form of a flared disk, whose 
outer rim simulates the warm photosphere. Material located 
outside of the disk's rim is in the ionization shadow and thus 
represents the neutral zone (Fig.~9). 

\subsection{Basic ionization structure}

The results we inferred above imply a significant 
difference in the ionization structure of the circumbinary 
environment of AX~Per between its quiescent and active phase. 
The corresponding sketch is shown in Fig.~9. It can be 
described as follows. 

During the quiescent phase, the steady-state STB \citep[][]{stb} 
model is sufficient to explain the radio emission from 
symbiotic stars and the wave-like orbitally related variation 
in the optical LCs \citep[][]{sk01}. 
However, to explain the shift of light minima to a position 
preceding the spectroscopic conjunction (Fig.~2, right panel), 
a more sophisticated model of the density distribution in the 
binary has to be considered. 
Hydrodynamical calculations of the structure of stellar winds 
in symbiotic stars that include effects of the orbital motion 
suggest a higher density region, twisted from the preceding 
front of the hot star motion to that of the giant 
\cite[e.g.][]{walder95,bis+06}. As a result, the light minima 
occur before the inferior conjunction of the giant 
\cite[see also Fig.~13 of][]{sk98}. In Fig.~9 we plotted a scheme 
of this region as calculated by \cite{bis+06} for a mass-loss 
rate from the giant and the hot star of $2\times 10^{-7}$ and 
$1.2\times 10^{-7}$\myr, respectively. 

During the active phase, the neutral hydrogen zone has the form 
of a flared disk encompassing the hot star. Its optically 
thick rim represents the warm pseudophotosphere. 
The ionized region fills in the remainder of the space 
above/below the neutral zone, which is ionized by the hot star. 
It can be associated with the enhanced stellar wind from 
the hot star, whose emission is subject to eclipse. 
The proposed ionization structure is consistent with the 
UV/near-IR model SED during active phases of symbiotic binaries 
with high orbital inclination \citep[see Fig.~27 in][]{sk05} 
and that of the bipolar stellar wind from the hot star used 
to model the broad \ha\ wings \citep[][]{sk06}. 
The formation mechanism of the neutral disk-like zone is not 
understood yet. However, first results based on the theory 
of the wind compression model \citep[][]{bjorkcass93} support 
the findings of this paper \citep[][]{c+s11}. 

\section{Conclusions}

We have investigated a new active phase of the eclipsing 
symbiotic binary AX~Per, which began during 2007. 
Our main conclusions may be formulated as follows. 

(i) 
AX~Per entered a new active phase from $\sim$2007.5 
(Sect.~3.1.1). After 10 orbital cycles ($\sim 18.6$ years) 
we again measured a deep narrow minimum in the LC, which was 
caused by the eclipse of the hot component by its giant companion. 
We improved the ephemeris of eclipses (Eq.~(1)) and found that 
their timing is identical (within uncertainties) with that of 
the inferior conjunction of the giant. We determined the radius 
of the giant as 
   $R_{\rm g} = 115 \pm 2$\ro\ 
and that of the eclipsed object, 
   $R_{\rm e} = 27 \pm 2$\ro\ 
(Sect.~3.1.2). 

(ii)
The new active phase was connected with a significant enhancement 
of the hot star wind. From quiescence to activity, the mass-loss 
rate increased from 
$\sim 9\times 10^{-8}$ to $\sim 2\times 10^{-6}$\myr, 
respectively (Sect.~3.2.2, Fig.~6, Table~7). 
The mean electron concentration within the eclipsed He$^{++}$ 
zone ($R_{\rm HeII} \sim 50$\ro\ above the hot star), 
$\sim 10^{10}$\cmt, agrees well with that given 
independently by the mass loss rate (Sects.~3.2.1 and 3.2.2). 
The hot star wind creates also a fraction of the [\ion{O}{iii}] 
zone. Depending on the mass-loss rate, the nebular 
lines can be subject to partial eclipse (1994 eclipse) or 
not (2009 eclipse; Sect.~ 4.1.1). 

(iii) 
We revealed a significant change of the ionization structure 
in the binary with respect to the quiescent phase (Sect.~4, Fig.~9). 
During the 2007-10 active phase we identified a warm stellar 
source located at the hot star equator (Sect.~3.2.3, Fig.~8). 
The source had the form of a flared disk, whose outer rim simulated 
the warm photosphere. It blocks out the ionizing radiation from 
the hot star, and consequently the material outside of the disk's 
rim is mostly in neutral form (Sect.~4.2). 
The ionized region is located above/below the neutral zone 
and can be associated  with the enhanced hot star wind. 
The formation of the disk-like neutral zone and the increase of 
the nebular emission during the active phase was connected with 
the enhanced wind from the hot star (Tables~7 and 8). 
It is probable that this connection represents a common origin 
of the warm pseudophotospheres that are observed during the 
active phases of symbiotic binaries. 

\begin{acknowledgements}
The authors would like to thank Theodor Pribulla for the acquisition 
of spectra at the David Dunlap Observatory. We also would like to 
thank S. Bacci, M. Marinelli, L. Ghirotto and A. Milani for assistance 
with some of the photometric and spectroscopic observations. 
We acknowledge with thanks the variable star observations from
the AAVSO International Database contributed by observers
worldwide and used in this research. 
The authors are grateful to the referee, Roberto Viotti, for detailed 
commenting on the manuscript and constructive suggestions. 
This research was in part supported by a grant of the Slovak  
Academy of Sciences, VEGA No.~2/0038/10. 
\end{acknowledgements}

\begin{thebibliography}{}
%
\bibliographystyle{aa}
%
\bibitem[Belyakina (1979)]{bel79}
         Belyakina, T. S. 1979, 
         Izv. Krym. Astrophys. Obs., 59, 133
%
\bibitem[Belyakina (1991)]{bel91}
         Belyakina, T. S. 1991, Bull. CrAO, 83, 104 
%
\bibitem[Bisikalo et al. (2006)]{bis+06}
         Bisikalo, D. V., Boyarchuk, A. A., Kilpio, E. Yu., 
         Tomov, N. A., \& Tomova, M. T. 
         2006, Astron. Reports, 50, 722
%
\bibitem[Bjorkman \& Cassinelli (1993)]{bjorkcass93}
         Bjorkman, J. E., Cassinelli, J.P. 1993, ApJ, 409, 429
%
\bibitem[Brown \& Mathews (1970)]{b+m70}
         Brown, R. L., \& Mathews, W. G.
         1970, ApJ, 160, 939
%
\bibitem[Carikov\'a \& Skopal (2010)]{c+s10}
         Carikov\'a, Z. \& Skopal, A. 2010, \na, 15, 637
%
\bibitem[Carikov\'a \& Skopal (2011)]{c+s11}
         Carikov\'a, Z. \& Skopal, A. 2011, in ASP Conf. Ser., 
         Evolution of Compact Binaries, 
         ed. L. Schmidtobreick, M. R. Schreiber, C. Tappert, 
         (San Francisco, CA: ASP), in press 
         (arXiv:1106.2420)
%
\bibitem[Castor et al. (1975)]{cak}
         Castor, J. I., Abbott, D. C., \& Klein R. I.
         1975, ApJ, 195, 157
%
\bibitem[Corradi et al. (2003)]{cmm03}
         Corradi, R. L. M., Mikolajewska, J., \& Mahoney, T. J.
         2003, Symbiotic Stars Probing Stellar Evolution, 
         ASP Conf. Ser. 303 (San Francisco: ASP)
%
\bibitem[Eriksson et al. (2004)]{eriksson}
         Eriksson, M., Johansson, S., \& Wahlgren, G. M. 
         2004, A\&A, 422, 987 
%
\bibitem[Fekel et al. (2000)]{fekel+00}
         Fekel, F. C., Hinkle, K. H., Joyce, R. R., 
         \& Skrutskie, M. F. 2000, AJ, 120, 3255
%
\bibitem[Fluks et al. (1994)]{fluks}
         Fluks, M. A., Plez, B., The, P. S., de Winter, D., 
         Westerlund, B. E., \& Steenman, H. C. 
         1994, A\&AS, 105, 311
%
\bibitem[Friedjung \& Viotti (1982)]{f+v82}
         Friedjung, M., \& Viotti, R. F. 1982, 
         The nature of symbiotic stars, IAU Coll. 70, 
         (Dordrecht, D. Reidel Publishing Co.)
%
\bibitem[Fujimoto (1982)]{fuji82}
         Fujimoto, M. Y. 1982, ApJ, 257, 767
%
\bibitem[Gurzadyan (1997)]{gurzadyan}
         Gurzadyan G. A., 1997, The Physics and Dynamics
         of Planetary Nebulae. Springer-Verlag, Berlin
%
\bibitem[Henden \& Munari (2006)]{h+m06}
         Henden, A. A., \& Munari, U. 
         2006, A\&A, 458, 339
%
\bibitem[Hummer \& Storey (1987)]{h+s87}
         Hummer, D. G. \& Storey, P. J. 1987, MNRAS, 224, 801
%
\bibitem[Iijima (1988)]{iijima88}
         Iijima, T. 1988, Ap\&SS, 150, 235
%
\bibitem[Ivison et al. (1993)]{ivison+93}
         Ivison, R. J., Bode, M. F., Evans, A., Skopal, A., 
         \& Meaburn, J. 1993, MNRAS, 264, 875
%
\bibitem[Kenny \& Taylor (2005)]{kenny05}
         Kenny, H. T., \& Taylor, A. R. 2005, ApJ, 619,527
%
\bibitem[Kenyon \& Webbink (1984)]{kw84}
         Kenyon, S. J., \& Webbink, R. F. 1984, ApJ, 279, 252
%
\bibitem[Mikolajewska \& Kenyon (1992)]{mk92}
         Mikolajewska, J. \& Kenyon, S. J. 1992, AJ, 103, 579
%
\bibitem[Munari \& Zwitter (2002)]{m+z02}
         Munari, U., Zwitter, T. 2002, A\&A, 383, 188
%
\bibitem[Munari et al. (2005)]{munari+05}
         Munari, U., Sordo, R., Castelli, F., \& Zwitter, T. 
         2005, A\&A, 442, 1127
%
\bibitem[Munari et al. (2009a)]{mu+09a}
         Munari, U., Siviero, A., Dallaporta, S., et al. 
         2009a, CBET No. 1757
%
\bibitem[Munari et al. (2009b)]{mu+09b}
         Munari, U., Siviero, A., Ochner, P., et al.
         2009b, PASP, 121, 1070
%
\bibitem[Munari et al. (2010)]{mu+10}
         Munari, U., Siviero, A., Corradi, R. L. M., Valisa, P., 
         Cherini, G., Castellani, F., \& Dallaporta, S.
         2010, CBET No. 2555
%
\bibitem[M\"urset \& Schmid(1999)]{m+s99}
         M\"urset, U., \& Schmid, H. M. 
         1999, A\&AS, 137, 473 
%
\bibitem[Nussbaumer \& Osterbrock (1970)]{no70}
         Nussbaumer, H., \& Osterbrock, D. E. 
         1970, ApJ, 161, 811
%
\bibitem[Nussbaumer \& Vogel (1987)]{nv87}
         Nussbaumer, H., \& Vogel, M. 1987, A\&A, 182, 51
%
\bibitem[Nussbaumer et al. (1995)]{nsv95}
         Nussbaumer, H., Schmutz, W., \& Vogel, M. 
         1995, A\&A, 293,13
%
\bibitem[Paczy\'nski \& \.{Z}ytkow (1978)]{paczyt78}
         Pac\'zynski, B., \& \.{Z}ytkow, A. N. 
         1978, ApJ, 222, 604
%
\bibitem[Paczy\'nski \& Rudak (1980)]{pacrud80}
         Paczy\'nski, B., \& Rudak, R. 1980, A\&A, 82, 349
%
\bibitem[Seaquist, Taylor, \& Button (1984)]{stb}
         Seaquist, E. R., Taylor, A. R., \& Button, S. 
         1984, ApJ, 284, 202
%
\bibitem[Siviero et al. (2009)]{siv+09}
         Siviero, A., Munari, U., Dallaporta, S., et al. 
         2009, MNRAS, 399, 2139
%
\bibitem[Skopal (1991)]{sk91}
         Skopal, A. 1991, IBVS No. 3603
%
\bibitem[Skopal (1994)]{sk94}
         Skopal, A. 1994, A\&A, 286, 453
%
\bibitem[Skopal (1998)]{sk98}
         Skopal, A. 1998, A\&A, 338, 599
%
\bibitem[Skopal (2000)]{sk00}
         Skopal, A. 2000, CoSka, 30, 21
%
\bibitem[Skopal (2001)]{sk01}
         Skopal, A. 2001, A\&A, 366, 157
%
\bibitem[Skopal (2005)]{sk05}
         Skopal, A. 2005, A\&A, 440, 995
%
\bibitem[Skopal (2006)]{sk06}
         Skopal, A. 2006, A\&A, 457, 1003
%
\bibitem[Skopal (2007)]{sk07}
         Skopal, A. 2007, \na, 12, 597
%
\bibitem[Skopal et al. (2001)]{sk+01}
         Skopal, A., Teodorani, M., Errico, L., Vittone, A. A., 
         Ikeda, Y., \& Tamura, S. 2001, A\&A, 367, 199
%
\bibitem[Skopal et al. (2006)]{sk+06}
         Skopal, A., Vittone, A. A., Errico, L., Otsuka, M., 
         Tamura, S., Wolf, M., \& Elkin, V. G. 
         2006, A\&A, 453, 279
%
\bibitem[Skopal et al. (2009)]{sk+09}
         Skopal, A., Seker\'a\v{s}, M., González-Riestra, R., 
         \& Viotti, R. F. 2009, A\&A, 507, 1531
%
\bibitem[Tamura et al. (2003)]{tamura+03}
         Tamura, S., Otsuka, M., Skopal, A., Pribulla, T., 
         \& Va\v{n}ko, M. 2003, in ASSL 298, Stellar 
         astrophysics - a tribute to Helmut A. Abt. Sixth Pacific 
         Rim Conference, ed. K.S. Cheng, K.C. Leung and T.P. Li 
         (Dordrecht, Kluwer Academic Publishers) p. 213
%
\bibitem[Tutukov \& Yangelson (1976)]{tutyan76}
         Tutukov, A. V., \& Yangelson, L. R. 
         1976, Astrofizika, 12, 521
%
\bibitem[Vogel (1993)]{vogel93}
         Vogel, M. 1993, A\&A, 274, L21
%
\bibitem[Walder (1995)]{walder95}
         Walder, R. 1995, in IAU Symp. 163, Wolf-Rayet Stars: 
         Binaries, Colliding Winds, Evolution, 
         ed. K.A. van der Hucht \& P.M. Williams, 
         (Dordrecht, Kluwer Academic Publishers) p. 420
%
\bibitem[Wallerstein et al. (1984)]{wall+84}
         Wallerstein, G., Willson, L. A., Salzer, J. \& Brugel 
         1984, A\&A, 133, 137
%
\end{thebibliography}
\end{document}